\documentclass[11pt]{article}
\usepackage{times}
\usepackage{geometry}
\usepackage[utf8]{inputenc}
\usepackage{enumitem,amssymb}
\usepackage{ragged2e}
\usepackage{soul}
\usepackage{multirow}
\newlist{thematic}{itemize}{8}
\setlist[thematic]{label=$\square$}
\usepackage{pifont}

\newcommand{\mission}{Voyage 2050 Backlight mission}
\newcommand{\Planck}{{\it Planck}}
\newcommand{\Euclid}{{\it Euclid}}
\newcommand{\LiteBIRD}{{\it LiteBIRD}}
\newcommand{\eROSITA}{{\it eROSITA}}
\newcommand{\Backlight}{{\sc Backlight}}

\usepackage[format=plain, font={footnotesize}, justification=centerlast, labelfont={footnotesize, bf}]{caption}
\usepackage{xspace}

\usepackage{ulem}
\usepackage{jheppub}   
\usepackage[dvipsnames,svgnames,x11names]{xcolor}

\usepackage{sectsty}
\sectionfont{\fontsize{15}{15}\selectfont}

\usepackage{sidecap}
\sidecaptionvpos{figure}{c}

\definecolor{DarkGreen}{rgb}{0.0, 0.3, 0.0}
\definecolor{purple}{rgb}{0.5, 0.0, 0.5}
\definecolor{red}{rgb}{1, 0.0, 0.0}
\definecolor{green}{rgb}{0, 1.0, 0.0}

\def\Mpc{{\rm Mpc}}

\def\3he{$^3{\rm He}$}

\def\lsim{\mathrel{\lower2.5pt\vbox{\lineskip=0pt\baselineskip=0pt
           \hbox{$<$}\hbox{$\sim$}}}}

\def\gsim{\mathrel{\lower2.5pt\vbox{\lineskip=0pt\baselineskip=0pt
           \hbox{$>$}\hbox{$\sim$}}}}

\newcommand{\eeq}{\end{equation}}

\usepackage{eso-pic}
\newcommand\BackgroundPic{%
\put(0,0){%
\parbox[b][\paperheight]{\paperwidth}{%
\vfill
\centering
\vspace{10cm}
\includegraphics[width=0.94\paperwidth,height=\paperheight,%
keepaspectratio]{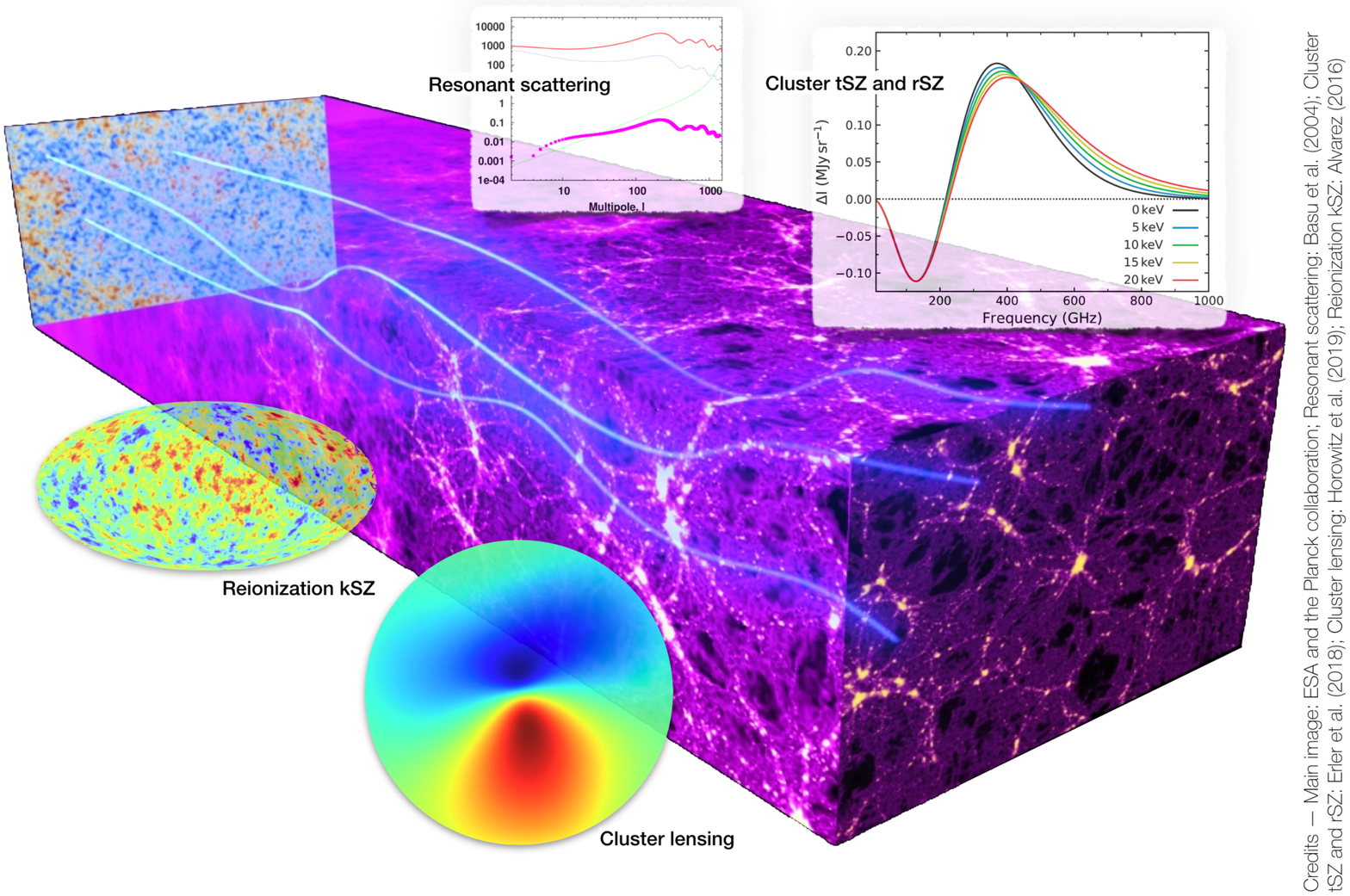}%
\vfill
}}}
\begin{document}

\AddToShipoutPicture*{\BackgroundPic}

\raggedright
\Large
\textbf{ESA Voyage 2050 Science White Paper} \linebreak
%\vspace{0.2cm}

\thispagestyle{empty}

\huge
A Space Mission to Map the Entire Observable \\ Universe using the CMB as a Backlight

\normalsize
\vspace{1cm}

%%%%%%%%%%%%%%%%%%%%%%%%%%%%
\textbf{Corresponding Author:} 
\smallskip

Name: Kaustuv Basu	
 \linebreak						
Institution: Argelander-Institut f\"ur Astronomie, Universit\"at Bonn, D-53121 Bonn, Germany 
 \linebreak
Email: kbasu@astro.uni-bonn.de, ~
% \linebreak
Phone: $+$49 228 735 658 
 \linebreak

\textbf{\small{Co-lead Authors:}} 
\smallskip

Mathieu Remazeilles$^{1}$ \textit{(proposal writing coordinator)}, 
~Jean-Baptiste Melin$^{2}$  
\vspace{2mm}

\noindent
{\scriptsize
$^{1}$ Jodrell Bank Centre for Astrophysics, Dept. of Physics \& Astronomy, The University of Manchester, Manchester M13 9PL, UK
\\[-1mm]
$^{2}$ IRFU, CEA, Universit{\'e} Paris-Saclay, F-91191 Gif-sur-Yvette, France
}

\newpage  
\thispagestyle{empty}
%\setcounter{page}{0}
%\vspace{-2mm}
\begin{minipage}[l]{0.94\columnwidth}

\textbf{Co-authors:} 
David Alonso$^{3,4}$, % 3
James G. Bartlett$^{5,6}$, % 4
Nicholas Battaglia$^{7}$, %6
Jens Chluba$^{1}$, % 7
Eugene Churazov$^{8,9}$, % 8
Jacques Delabrouille$^{2,5}$, % 9
Jens Erler$^{10}$, % 10
Simone Ferraro$^{11,12}$, % 11
Carlos Hern\'andez-Monteagudo$^{13}$, % 12
J.~Colin Hill$^{14,15}$, % 13
Selim~C.~Hotinli$^{16}$, % 14
Ildar Khabibullin$^{8,9}$, % 15
Mathew Madhavacheril$^{17}$, % 16
Tony Mroczkowski$^{18}$, % 17
Daisuke Nagai$^{19}$, % 18
Srinivasan Raghunathan$^{20}$, % 19
Jose Alberto Rubino Martin$^{21,22}$, % 20
Jack Sayers$^{23}$, % 21
Douglas Scott$^{24}$, % 22
Naonori Sugiyama$^{25}$, % 23
Rashid Sunyaev$^{8,9,14}$, % 24
\'I\~{n}igo Zubeldia$^{26,27}$. % 25
  \linebreak 
  
\vspace{-1mm}
\noindent
{\scriptsize
  
% Alonso  
$^{3}$ University of Oxford, Denys Wilkinson Building, Keble Road, Oxford, OX1 3RH, UK 
\\
%  Alonso
$^4$ School of Physics and Astronomy, Cardiff University, The Parade, Cardiff, CF24 3AA, UK 4 
\\
%
% Delabrouille, Bartlett
$^{5}$ Laboratoire Astroparticule et Cosmologie (APC), CNRS/IN2P3, Universit\'e Paris Diderot, 75205 Paris Cedex 13, France 
\\
%
% Bartlett
$^{6}$ Jet Propulsion Laboratory, California Institute of Technology, 4800 Oak Grove Drive, Pasadena, CA, USA 91109 
\\
% Battaglia
$^{7}$ Department of Astronomy, Cornell University, Ithaca, NY 14853, USA
\\
%
% Chluba
%$^{8}$ Jodrell Bank Centre for Astrophysics, Dept. of Physics \& Astronomy, The University of Manchester, Manchester M13 9PL, UK
%\\
%
% Churazov, Khabibullin, Sunyaev
$^{8}$ Max-Planck-Institut f\"ur Astrophysik, Karl-Schwarzschild Str. 1, 85741 Garching, Germany
\\
%
% Churazov, Khabibullin, Sunyaev
$^{9}$ Space Research Institute (IKI), Profsoyuznaya 84/32, Moscow 117997, Russia
\\
%
% Delabrouille, Melin
%$^{10}$ CEA Saclay DSM/Irfu, Universit\'e Paris Saclay, 91191 Gif-sur-Yvette, France 
%\\
%
% Basu, Erler
$^{10}$ Argelander-Institut f\"ur Astronomie, Universit\"at Bonn, Auf dem H\"ugel 71, D-53121 Bonn, Germany
\\
%
% Ferraro
$^{11}$ Berkeley Center for Cosmological Physics, University of California, Berkeley, CA 94720, USA
\\
%
% Ferraro
$^{12}$ Lawrence Berkeley National Laboratory, One Cyclotron Road, Berkeley, CA 94720, USA
\\
%
% Hernandez-Monteagudo
$^{13}$ Centro de Estudios de F\'isica del Cosmos de Arag\'on (CEFCA), Plaza San Juan, 1, planta 2, E-44001, Teruel, Spain
\\
%
% Hill, Sunyaev
$^{14}$ Institute for Advanced Study, Princeton, NJ 08540, USA
\\
%
% Hill
$^{15}$ Center for Computational Astrophysics, Flatiron Institute, 162 5th Avenue, New York, NY 10010, USA
\\
% Hotinli
$^{16}$ Imperial College London, Blackett Laboratory, Prince Consort Road, London SW7 2AZ, UK
\\
%
% Madhavacheril
$^{17}$ Perimeter Institute for Theoretical Physics, Waterloo, ON N2L 2Y5, Canada
\\
% 
% Mroczkowski
$^{18}$ European Southern Observatory, Karl-Schwarzschild-Strasse 2, Garching D-85748, Germany
\\
%
% Nagai
$^{19}$ Department of Physics \& Department of Astronomy, Yale University, New Haven, CT 06520, USA
\\
%
% Raghunathan
$^{20}$ Department of Physics and Astronomy, University of California, Los Angeles, CA 90095, USA
\\
%
% Rubino-Martin
$^{21}$ Departamento de Astrof\'{\i}sica, Universidad de La Laguna, E-38206 La Laguna, Tenerife, Spain
\\
$^{22}$ Instituto de Astrof\'{\i}sica de Canarias, C/ V\'ia L\'actea 39020 La Laguna (Tenerife), Spain
\\
%
% Sayers
$^{23}$ California Institute of Technology, 1200 E. California Boulevard, MC 367-17, Pasadena, CA 91125, USA
\\
%
% Scott
$^{24}$ Department of Physics and Astronomy, University of British Columbia, 6224 Agricultural Road, Vancouver, V6T 1Z1, Canada
\\
%
% Sugiyama
$^{25}$ National Astronomical Observatory of Japan, Mitaka, Tokyo 181-8588, Japan
\\
%
% Zubeldia
$^{26}$ Institute of Astronomy, Madingley Road, Cambridge CB3 0HA, UK
\\
% Zubeldia
$^{27}$ Kavli Institute for Cosmology Cambridge, Madingley Road, Cambridge CB3 0HA, UK

 } 
\end{minipage}

\justify 
%\textbf{Abstract:} 

%\newpage
\tableofcontents
\newpage
\setcounter{page}{1}

\pagebreak

%%%%%%%%%%%%%%%%%%%%%%%%%%%%%%%%%%%%%%%%%%%%%%%%%%%%%%%%%%%%%%%%%
\vspace{2mm}
\section*{Executive Summary}
\label{sec:execsum}

Cosmology came of age over the past two decades with the establishment of the standard cosmological model ($\Lambda$CDM) in the 1990s, followed by rapid progress up to the present in the determination of its basic parameter values.
The European Space Agency's (ESA) \Planck\ mission played a crucial role in this scientific success by measuring  cosmological parameters to an unprecedented precision of better than one percent. Flagship experiments, such as the Large Synoptic Survey Telescope (LSST) and ESA's \Euclid\ mission, will herald the decade of the 2020s by characterizing dark energy across cosmic time and testing for deviations from General Relativity. By the late 2020s, the \LiteBIRD\ mission and the CMB-S4 experiment may reveal the secrets of the inflationary phase at the very beginning of cosmic time. Looking even farther, to the middle of the 2030s, we present in this white paper what we expect to be some remaining key questions in cosmology and propose a mission concept to address them.

By the middle of the 2030s, we should attain even more precise measurements of the fundamental parameters of the standard cosmological model (including the neutrino mass scale), an initial characterization of the dark-energy equation-of-state over time, and the first precision tests of General Relativity on cosmic scales.  We also expect that understanding dark energy and searches for modifications to General Relativity will demand higher precision and accuracy, and that a complete picture of structure formation and evolution will still elude us, leaving the following fundamental questions.

\begin{itemize}
\item What are the natures of dark energy and dark matter, and how is dark matter distributed? Are there deviations from General Relativity, and on what scales?
\item What is the relationship between dark matter and ordinary baryonic matter?  What are their relative distributions in the Universe and how do they interact, from galactic to cosmic scales? 
\item How do the baryons in the Universe evolve from primordial atomic gas to stars within galaxies? How is feedback so finely tuned to allow only 10\% to form stars, and what is the nature and distribution of the gas containing the other 90\%?
\end{itemize}

We propose to answer these questions using the \textbf{\textit{cosmic microwave background (CMB) as a ``backlight''}}, illuminating the entire observable Universe from the epoch of its emission at recombination until today. Structures along the line-of-sight imprint small distortions in the spatial structure and frequency spectrum of the CMB which trace the baryonic and dark-matter distributions and velocities.  Our proposed ``Backlight Mission'' would use these signals to achieve, for the first time, a complete census of the total mass, gas and stellar contents of the Universe and their evolution from the earliest times.

Examples of this capability include: the detection of all massive bound structures ($M>5 \times 10^{13} {\rm M}_\odot$) in the observable Universe; routine measurement of CMB halo lensing and the kinetic Sunyaev-Zeldovich (SZ) effect; measurement of the relativistic SZ effect in individual halos; the first detection of the polarized SZ effects; and investigation of non-thermal SZ effects and resonant scattering of CMB photons. The diversity of this non-exhaustive list illustrates the richness of CMB backlight science.  It is this breadth that makes it a uniquely powerful and hitherto unexploited resource for completing our cosmic census.

%\fbox{\begin{minipage}[t][][b]{0.9\textwidth} %\hfill
%\justify
Our goals require an all-sky polarization survey in at least 20 channels over a frequency range from 50\,GHz to 1\,THz with a resolution better than 1.5$^\prime$ at 300\,GHz (goal 1$^\prime$) and an average sensitivity of a few 0.1$\,\mu$K-arcmin.\footnote{$\mu$K-arcmin is a common unit for characterizing CMB map noise, assuming the noise properties are Gaussian. It is defined as the rms of the CMB temperature fluctuations within a map created with pixels that each subtend a solid angle of 1 square arcmin.} The need to resolve individual structures (filaments and halos down to a mass $10^{14} {\rm M}_\odot$ out to $z=1$) sets the  angular resolution, which in turn calls for at least a 3--4-m class telescope and, preferably, a 4--6-m class to attain the 1$^\prime$ goal. This spectral coverage and sensitivity can only be achieved from space, and the telescope size will require an L-class mission.
%\end{minipage}}

%\vspace{5mm}
While our Backlight mission, to which we also refer simply as \Backlight, can reach its science goals as described herein without the need of additional, external millimeter data, we note that the resolution of 1.5$^\prime$ at 300\,GHz matches the resolution of the future ground-based CMB-S4 experiment. CMB-S4 will operate in atmospheric windows at the same resolution as \Backlight\ at frequencies below 300\,GHz. The two experiments would therefore enjoy a powerful synergy for further exploration of many of the science cases that we now describe.

%%%%%%%%%%%%%%%%%%%%%%%%%%%%%%%%%%%%%%%%%%%%%%%%%%%%%%%%%%%%%%%%%
\vspace{2mm}
\section{Introduction: Fundamental Science Questions and Methods}

The CMB is the oldest source of light in the Universe, emanating from the last scattering surface, about 380,000 years after the Big Bang (or $z\approx1100$) when neutral atoms first formed, and cooling with the expansion until we observe it today as a nearly perfect blackbody spectrum with a temperature of $2.7255(6)\,{\rm K}$.
Primary CMB anisotropies  carry invaluable information about the physics of the early Universe prior to last scattering.  More importantly for our objectives, the CMB also presents a bright screen, or {\it backlight}, against which cosmic structure has emerged and evolved since $z\approx1100$.  The gravitational field of these structures deviated the path of the photons, while scattering by ionized gas in the cosmic web distorted their energy spectrum, imprinting the CMB with numerous telltale secondary anisotropies. 

In this paper, we develop the case for \emph{using the CMB as a backlight} to probe cosmic structures, such as cosmic filaments, galaxy clusters, groups and galaxy-sized halos. Although we anticipate tremendous progress in cosmology by the mid-2030s, many fundamental questions about structure formation and evolution can be expected to remain unresolved: %\comm{JC: just replace 'unanswered' by 'open' or even just remove the word?}
\begin{enumerate}
\item The Dark Sector: What exactly are the dark energy and dark matter, and how are they distributed? Are there modifications to General Relativity on large scales?
\item The Cosmic Web: How exactly do baryons populate dark-matter halos? What are their relative distributions in the Universe and how do they interact, from sub-galactic to Hubble scales? 
\item State of the Baryons: How do the baryons in the Universe evolve from primordial atomic gas to stars within galaxies? How do feedback processes modify this, allowing only 10\% to form stars? In what form are the baryons in the remaining 90\%?
\end{enumerate}

A Voyage 2050 space mission optimized for ``backlight science" can definitively answer these fundamental questions.  The  tools of this science are the Sunyaev-Zeldovich (SZ) effects (thermal, kinetic, relativistic, polarized, and non-thermal) and gravitational lensing of the CMB.  Section~\ref{sec:scicase} presents example case studies in their application with the aim of answering these questions, and that when combined will give us the first complete census of all matter and its evolution in the Universe. 

Electrons in hot ionized gas, such as the intracluster medium (ICM) or circumgalactic medium (CGM), distort the frequency spectrum of the CMB by transfering energy to CMB photons through inverse Compton scattering.  This thermal SZ effect (tSZ)  was detected for the first time in the 1980s \cite{Birkinshaw1984}, but was routinely exploited only by the 2000s to build new, large galaxy cluster catalogues containing thousands of objects \cite{Staniszewski2009,Menanteau2010,PlanckESZ}. The tSZ effect is now established as a robust and powerful means of detecting galaxy clusters with a well characterized selection function, close to pure mass selection. 

The tSZ effect continues to hold enormous promise. The Advanced ACT~\cite{Henderson2016} and SPT-3G experiments~\cite{Benson2014} will detect of order of $10^4$ clusters over the next five years, and the CMB-S4 experiment \cite{CMBS42016} of order of $10^5$ clusters by around 2030, and \Backlight\ as proposed here and in associated papers \citep{Delabrouille2019WP, Chluba2019WP} would push the field to a comprehensive census of all massive structures in the observable Universe by detecting 1,000,000 clusters and groups (Sec.~\ref{sec:counts}). Such a complete  sample of structures would constrain key cosmological parameters independent from the primary CMB anisotropies, providing critical tests of the cosmological model. 

If the ionized gas has non-zero velocity relative to the CMB rest-frame, the scattering induces a Doppler shift in the photons to create a distortion known as the kinetic SZ (kSZ) effect, whose amplitude is directly sensitive to the radial velocity of the structure. Typically ten times smaller than the tSZ signal for clusters, initial detections of the kSZ effect has been more difficult \cite{Hand12}. Our Backlight mission will enable routine measurement of the kSZ to map the large-scale velocity field (see Sec.~\ref{sec:kSZveloc},\ref{sec:filaments},\ref{sec:feedback}). 

The polarized SZ effects (pSZ), although even smaller and not yet detected, would measure transverse velocities of the structures in the plane of the sky. 
In the most massive clusters, with temperatures high enough to populate relativistic electron energy states, the relativistic SZ (rSZ) effect subtly modifies the spectrum of the tSZ effect. Not yet detected for individual objects, the sensitivity and resolution of \Backlight\ would observe the rSZ effect and measure the temperature of gas in massive structures, offering a new proxy for determining cluster masses (see Sect.~\ref{sec:rSZ}).

The SZ effects trace {\it baryons} in their dominant form of ionized gas within large-scale structure. Gravitational lensing provides a way to probe the {\it total mass}, by measuring the deflection of CMB light rays by foreground structures. 
The lensing of CMB by halos was first detected in 2014 \cite{Madhavacheril2015, Baxter2015, PlanckClusterCosmo2016} with temperature and in 2019 \cite{raghunathan2019} with polarization data in cluster and group sized objects. CMB lensing will be the most powerful way to measure object masses with the \mission\ and has the potential of reducing the uncertainties in cluster masses to sub-percent levels out to high redshifts. The polarization channel will be particularly useful as it is less susceptible to the extragalactic foreground signals that are largely unpolarized (see Sec.~\ref{sec:lensing}, \ref{sec:fnl}).

Finally, in addition to a census of ionized gas and total mass, \Backlight\ will also build a {\it stellar} census by observing dust emission in galaxies, groups, and clusters across cosmic time, thanks to its high resolution and high frequency ($>500 \,{\rm GHz}$) coverage. This specific science case is detailed in the Voyage 2050 white paper~\cite{Delabrouille2019WP}.
\emph{This census of the {\it gas}, {\it total mass}, and {\it stellar content} of the entire observational Universe is the defining science goal for a \mission.}

%%%%%%%%%%%%%%%%%%%%%%%%%%%%%%%%%%%%%%%%%%%%%%%%%%%%%%%%%%%%%%%%%
\vspace{2mm}
\section{Detailed Science Case Studies}
\label{sec:scicase}
As examples, we describe a number of specific science case studies organized around our primary questions.  
%The last example in each section straddles the adjoining questions and bridges the goals in each. 
Some of these topics also straddle the adjoining science questions in each section and bridge their goals.

\subsection{The Dark Sector: Dark Energy, Dark Matter and Gravity}
\subsubsection{Galaxy cluster number counts from the tSZ effect} 
%\textcolor{red}{[J.-B. Melin, J. Bartlett]}
\label{sec:counts}

\begin{figure}[htbp!]
\centering
%\begin{center}
  \includegraphics[width=0.5\textwidth]{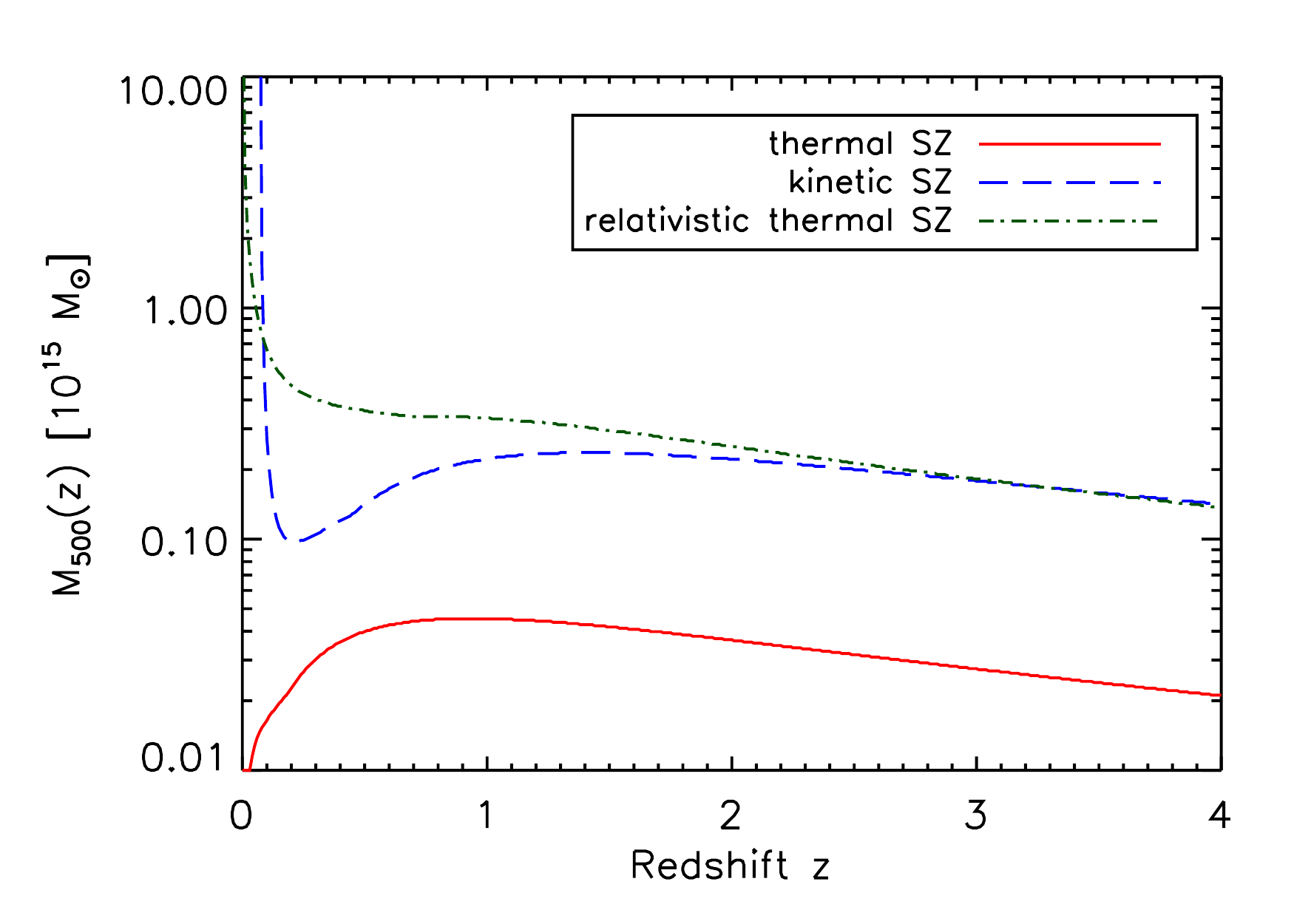}~\includegraphics[width=0.5\textwidth]{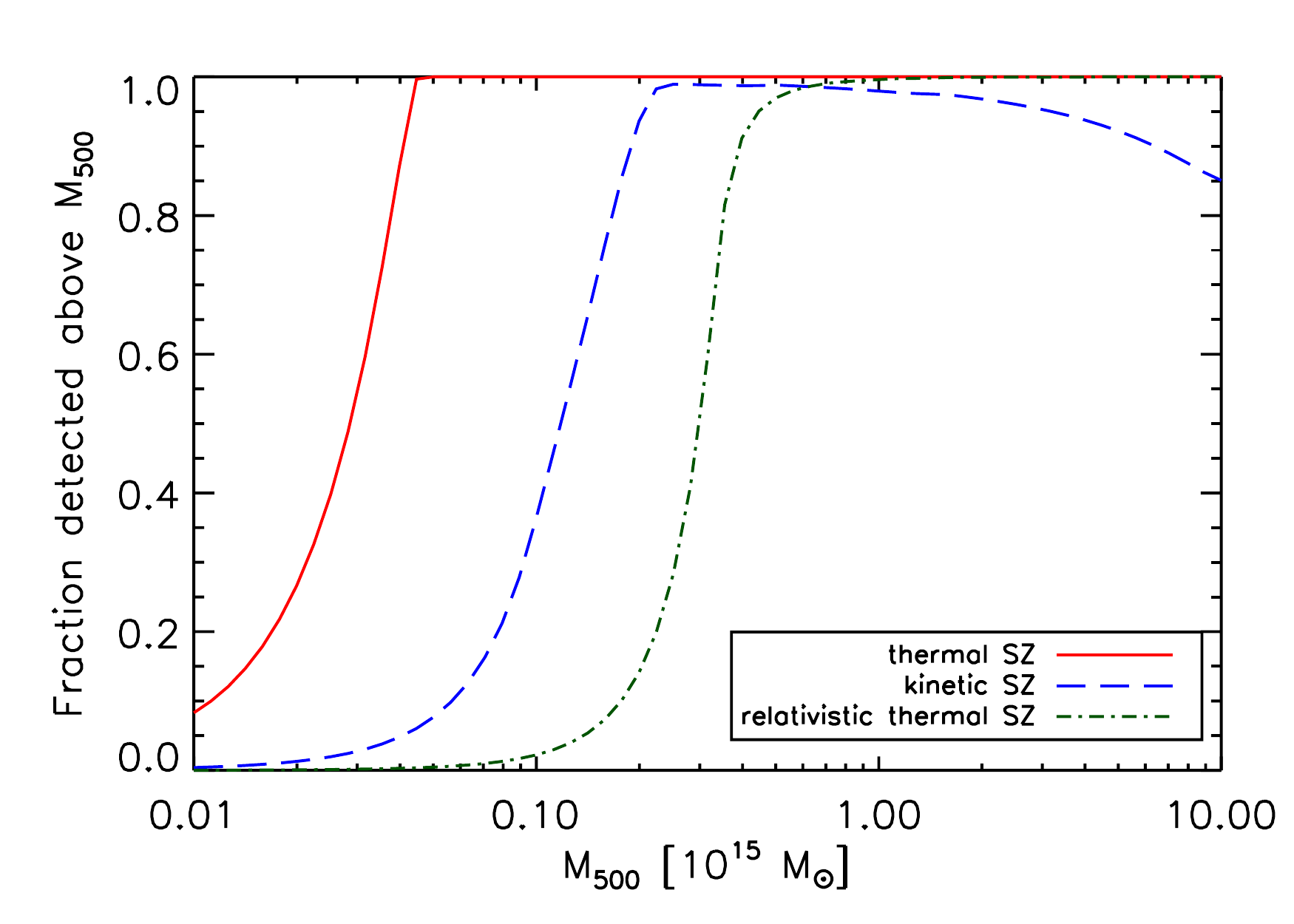}
%\end{center}
\caption{\small {\it Left:} Mass detection limit across redshifts for the thermal SZ effect (solid red line), the kinetic SZ effect (dashed blue line), and the relativistic SZ effect (dotted dashed green line) for a ${\rm S/N}>5$ threshold. {\it Right:} Fraction of clusters detected at ${\rm S/N}>5$. {\it All} the clusters in the Universe with mass above $5 \times 10^{13} {\rm M}_\odot$ are detected via their thermal SZ effect. Figures from Ref.~\cite{PRISM2014}. For comparison, the detection limit for CMB-S4 is expected to be around $10^{14} {\rm M}_\odot$, as shown in figure~79 of Ref.~\cite{S4DSR19}.}
\label{fig:masslim}
\end{figure}

Galaxy clusters form the nodes of the filamentary structure of the cosmic web. They contain a wealth of information for studying cosmology and structure formation. Number counts across cosmic time are very sensitive to cosmological parameters. Clusters are also ideal laboratories, as virialized objects, to study galaxy evolution. These facts motivated and still motivate large surveys for finding higher redshift and less massive clusters. A \mission\ will make a breakthrough by completing for the first time a full census of galaxy clusters in the observable Universe.  It would detect of order of $10^6$ clusters thanks to the tSZ effect~\cite{PRISM2014}. It would supersede by a factor of 100 the number of clusters that the ongoing SPT-3G experiment will detect, and by a factor of 10 the number of clusters provided by the recently launched {\it eROSITA} mission, the coming {\it Euclid\/} mission, and future CMB-S4 experiment. It will also detect thousands of clusters above $z=2$. This latter number depends on the evolution of the SZ flux-mass ($Y$--$M$) relation, which we have assumed to be self-similar. There is no indication to date for a break in self-similarity of this relation, either from observations or from simulations.
The predicted number counts correspond to a limiting mass detection threshold at ${\rm S/N}>5$ above $5 \times 10^{13} {\rm M}_\odot$ at all redshifts, as shown in the left panel of Fig.~\ref{fig:masslim}. This mass limit allows the detection of {\it all} clusters above this mass in the observable Universe, as shown by the solid red line in right panel of Fig.~\ref{fig:masslim}.
The kSZ/rSZ effect will be detected in individual clusters of mass greater than $2 \times 10^{14} {\rm M}_\odot$ / $5 \times 10^{14} {\rm M}_\odot$, respectively. These predictions are based on Ref.~\cite{PRISM2014} which assumed a sensitivity 2 times worse than the sensitivity we envisage for \Backlight. By the time \Backlight\ flies, many wide-area surveys such as {\it Euclid}, LSST, DESI, MSE, and SKA, will provide imaging and spectroscopy over a large fraction of the sky, which will help for identifying optical/radio counterparts and determining redshifts for the new sources detected by the mission.
The \mission\ would thus provide a unique opportunity to complete a full census of the structures in the Universe across all redshifts.
It would be advantageously complemented by CMB-S4, which will provide the lower frequency channels ($<220\,{\rm GHz}$) at the same angular resolution as the high frequency ($>220\,{\rm GHz}$) channels of \Backlight.

\vspace*{-3mm}
\subsubsection{Lensing calibration of cluster masses} 
\label{sec:lensing}

The hundreds of thousands of clusters expected to be detected by \Backlight\ provided in Sect.~\ref{sec:counts} will be crucial to constrain the cosmological parameters that influence the geometry and growth of structures in the Universe. 
The cluster abundance measurements when combined with independent geometrical CMB and BAO measurements can also greatly enhance the constraining power of the overall cosmological parameter results \citep{PlanckClusterCosmo2016, hasselfield13, deHaan:2016qvy, bocquet2019, zubeldia19}. 
An important step in achieving this, however, involves an accurate measurement of the cluster masses. 
The masses measured using SZ effect, optical richness, or X-rays are subject to biases owing to the assumptions in converting those survey observable to cluster masses. 

The most robust and unbiased technique for measuring cluster masses is via weak lensing measurements. To facilitate lensing measurements, a background light source is required which could either be galaxies or the CMB. At low redshifts, galaxy weak lensing will provide a higher S/N compared to CMB weak lensing measurements (see Fig.\,3 of \cite{Madhavacheril2017}). By contrast, since the CMB originates behind all of the clusters, lensing of the CMB by clusters is a highly promising tool for measuring masses of clusters above $z \ge 1$ \citep{lewis06}. At these redshifts, the rate of lensed background galaxies observed with high S/N drops significantly. 

The CMB-cluster lensing measurement can be performed with either CMB temperature or polarization anisotropies. 
The temperature-based lensing measurement has been detected by several experiments \cite[e.g.,][]{Madhavacheril2015, Baxter2015, PlanckClusterCosmo2016} and a first
detection of the polarized CMB-cluster lensing has also been recently made \citep{raghunathan2019}. 
For the current surveys, the mass constraints obtained from polarization-based measurements are much weaker than temperature, since the lensing amplitude is higher in temperature. 
However, the CMB temperature data are highly susceptible to contaminating foreground signals originating from the cluster (tSZ, kSZ) and other extragalactic sources \cite{Madhavacheril-Hill2018}. 
As a result, CMB polarization will be the primary channel for CMB-cluster lensing measurements from the upcoming low-noise surveys \citep{raghunathan2017} like CMB-S4, and will allow us to calibrate the mass scaling relations of other observables, like optical richness and SZ/X-ray flux.

\begin{figure}[htbp!]
\centering
%\begin{center}
  %\includegraphics[width=0.9\textwidth]{Figures/cluster_mass_constraints_cmb_lensing.pdf}
 \includegraphics[width=0.5\textwidth]{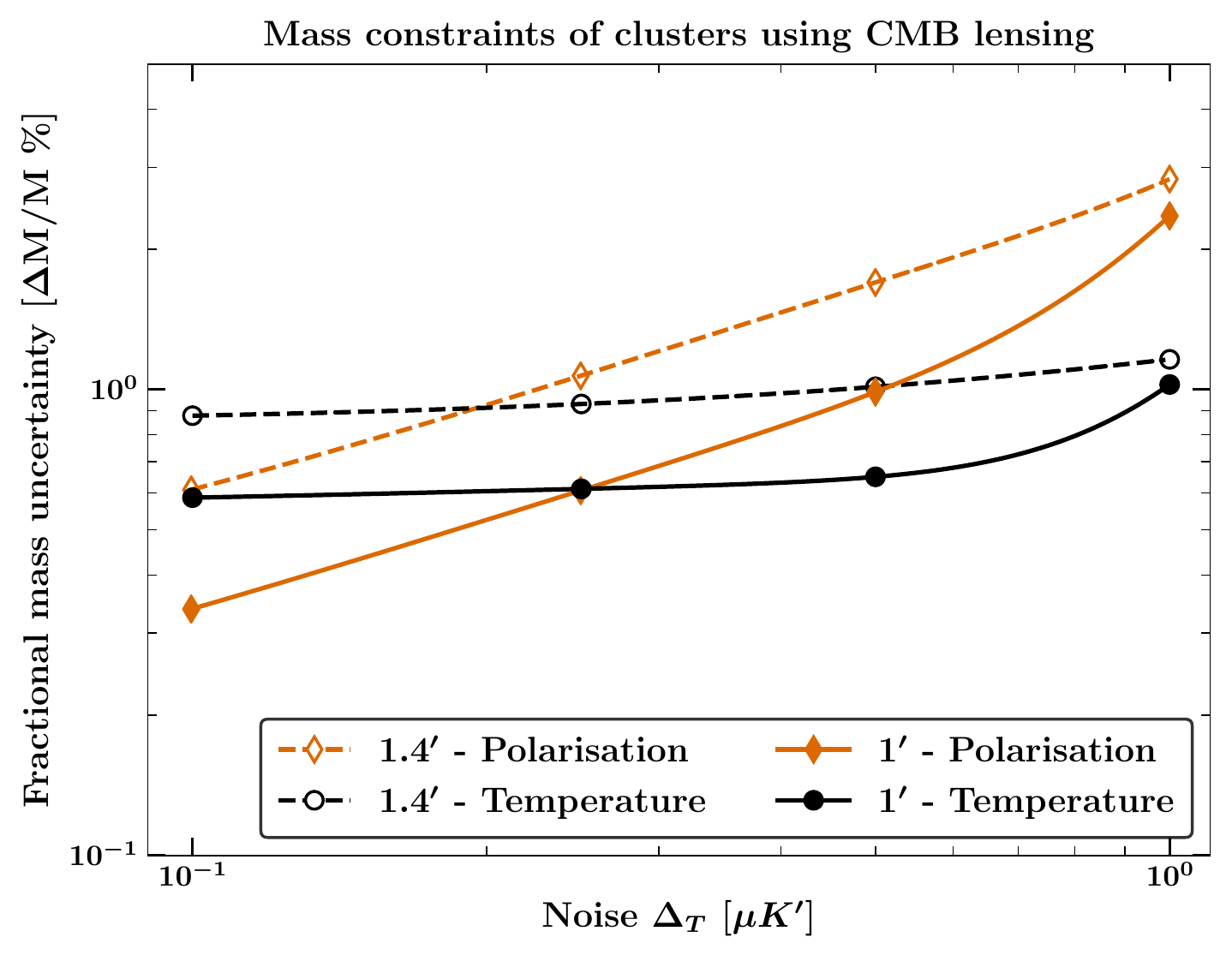}
%\end{center}
\caption{\small Mass uncertainties for a cluster sample containing 25,000 clusters derived using CMB-lensing measurements from temperature (black circles) and polarization (orange diamonds) data. The open and filled data points correspond to the constraints expected from a telescope with resolution $1^{\prime}$ (6-m aperture) and $1.4^{\prime}$ (4-m aperture) at 220\,GHz, respectively. 
The flattening of the constraints from CMB temperature data is due to the floor set by kSZ signals from unresolved halos that cannot be removed using data from multiple frequencies. 
}
\label{fig_cmb_cluster_lensing_mass_constraints}
\end{figure}

\Backlight\ is expected to return even deeper maps than CMB-S4, with exquisite control of the foreground contamination through unprecedented broad spectral coverage. 
However, while most of the frequency-dependent foregrounds can be cleaned by combining temperature data from multiple frequencies, the kSZ signal, which has the same blackbody spectrum as the CMB, cannot be removed by this technique. 
As a result, the kSZ signal sets a floor for the lensing measurements using CMB temperature data.
CMB polarization, on the other hand, is largely insensitive to the foregrounds and hence will be crucial for the lensing measurements with \Backlight.
We demonstrate this in Fig.~\ref{fig_cmb_cluster_lensing_mass_constraints}, which gives the uncertainty in the stacked mass of a cluster sample containing 25,000 clusters for different telescope aperture sizes (dashed lines represents an aperture size of 4\,m and the solid lines are for 6\,m). 
For simplicity, all clusters were placed at $z=0.7$ with a mass of $2 \times 10^{14}\  {\rm M}_{\odot}$.
CMB polarization, which is robust against foreground signals, can provide better mass constraints than temperature at noise levels lower than $\Delta_{T} = 0.25\ \mu$K-arcmin. 
Depending on the noise level, these measurements will be a factor of 2--10 better than those expected from the CMB-S4 survey \citep[see also Fig.~2 of Ref.][]{raghunathan2017} which is forecasted to have a map noise level of $\Delta_{T} = 2\ \mu$K-arcmin \citep{S4DSR19}.

\vspace*{-3mm}
\subsubsection{Cosmic velocity fields with the kSZ and moving lens effects} 
\label{sec:kSZveloc}

The relation between the density and velocity fields contains valuable information for cosmology, since it can be used to measure the growth rate of structure, which is strongly affected by the details of the late-time accelerated expansion and the properties of the field or modifications to the standard gravity theory that could source it. This is an area where current and future measurements of the kSZ effect could have a transformational impact. The kSZ effect is caused by the Doppler kick induced on CMB photons by moving free electrons, giving rise to thermal temperature fluctuations on the CMB that are sensitive to the gas radial peculiar momentum, $p_{\rm e}\propto \bar{n}_{\rm e} (1+\delta_{\rm e}) \mathbf{v}_{\rm e}\cdot \hat{\mathbf{n}}$, where $\bar{n}_{\rm e} (1+\delta_{\rm e})$ is the electron number density and $\mathbf{v}_{\rm e}$ their peculiar velocity (in units of $c$) at a given distance along the line of sight given by $\hat{\mathbf{n}}$).  More accurately, the temperature anisotropy caused by this effect along a line of sight $\hat{\bf n}$ is proportional to the radial projection of this quantity: $\Delta T_{\rm kSZ}(\hat{\bf n}) / T_{\rm CMB} =-\sigma_{\rm T}\int \frac{d\chi}{1+z}\bar{n}_{\rm e}(z)[1+\delta_{\rm e}(\chi\hat{\bf n})]\,{\bf v}_{\rm e}(\chi\hat{\bf n})\cdot\hat{\bf n}$,
%\begin{equation}
%  \frac{\Delta T_{\rm kSZ}(\hat{\bf n})}{T_{\rm CMB}}=-\sigma_T\int \frac{d\chi}{1+z}\bar{n}_e(z)[1+\delta_e(\chi\hat{\bf n})]\,{\bf v}_e(\chi\hat{\bf n})\cdot\hat{\bf n},
%\end{equation}
where $\sigma_{\rm T}$ is the Thomson scattering cross section and $\chi$ the comoving radial distance to redshift $z$.

Given the degeneracy between the electron number density and the radial velocity, a measurement of the kSZ momentum can be interpreted as an estimate of the cosmic peculiar velocity field when the amount of free electrons is known, or alternatively, be used as a {\it leptometer\/} when the peculiar velocity field has been constrained by other observations. 
%for instance, optimized reconstruction methods yielding peculiar velocity estimates in every point sampled by an input redshift galaxy survey. 
%As mentioned above, the most promising avenue to obtain 
%
As mentioned above, the cosmological application follows from measuring the growth rate of structure, $f\equiv d\log\delta/d\log a$, which connects the density and velocities as ${\bf v}_{\bf k}\propto f\,\frac{i{\bf k}}{k^2}\delta_{\bf k}$,
%\begin{equation}
%  {\bf v}_{\bf k}\propto f\,\frac{i{\bf k}}{k^2}\delta_{\bf k},
%\end{equation}
and quantifies how fast density inhomogeneities grow in time. A precise measurement of $f$, particularly at low redshifts, would be extremely powerful to constrain alternative dark energy models or neutrino masses \citep{DeDeo:2005yr,HernandezMonteagudo:2005ys,Bhattacharya:2007sk,Kosowsky:2009nc,Keisler:2012eg,Ma:2013taq,Mueller:2014nsa,Mueller:2014dba,Alonso:2016jpy}. 
The most promising sources for kSZ measurements are clusters of galaxies, which have minimal contamination from unbound gas along the same line of sight, and are easy targets to obtain alternative estimates of their mass through multi-wavelength observations (tSZ, X-ray, optical/infra-red).

\begin{table}
{\footnotesize
\begin{center}
\begin{tabular}{|c|c|c|c|}
\hline
Data combination            & Method   & Predicted S/N & Reference \\ \hline \hline
SPT-3G $\times$ DES      & pairwise kSZ & 18--30         & \citep{Keisler:2012eg}   \\ \hline
Adv.ACTPol $\times$ DESI & pairwise kSZ & 20--57         & \citep{Flender2016}  \\ \hline
Adv.ACTPol $\times$ SPHEREx & pairwise kSZ & 55 & \citep{Dore:2016tfs} \\ \hline
Adv.ACTPol $\times$ WISE & projected kSZ & 120 & \citep{Ferraro:2016ymw} \\ \hline
Adv.ACTPol $\times$ DESI & pairwise kSZ power spectrum & 30 & \citep{Sugiyama:2016rue} \\ \hline
CMB StageIV $\times$ DESI & pairwise kSZ power spectrum & 50--100 & \citep{Sugiyama:2016rue} \\ \hline
\end{tabular}
\end{center}
}
\caption{\label{table_forecasts} \small kSZ detection forecasts for future experiments from various published studies.}
\end{table}

Several methods have been put forward in the literature to extract information from the kSZ effect. The most widespread technique is the so-called {\sl mean pairwise velocity} (or pairwise momentum) estimator, which constrains the mean relative velocity between pairs of mass tracers (e.g. galaxy clusters) as a function of their separation \citep{1999ApJ...515L...1F,Hand12}. To date, the pairwise kSZ signal has been measured using CMB data from the Atacama Cosmology Telescope (ACT) with galaxy positions from the Baryon Oscillation Spectroscopic Survey (BOSS) \citep{Hand12,DeBernardis2017}, CMB data from {\it Planck\/} with galaxy positions from the Sloan Digital Sky Survey (SDSS) \citep{Ade:2015lza,Sugiyama2018}, and CMB data from the South Pole Telescope (SPT) with cluster positions from the Dark Energy Survey (DES) \citep{Soergel:2016mce}.
%
%In addition to these pairwise kSZ measurements, other methods have also been put forward. 
Another possibility is to detect the kSZ signal by stacking CMB patches around galaxy locations weighted by the local velocity field reconstructed from a spectroscopic galaxy survey \citep{Schaan:2015uaa,Alonso:2016jpy}. This method requires precise redshift information for a precise reconstruction of the 3D density and velocity fields. It is also possible to make use of 2D photometric surveys, by cross-correlating the galaxy positions with squared CMB temperature maps \citep{Hill:2016dta,Ferraro:2016ymw}. As shown in Ref.~\cite{Smith2018}, most of these methods can be encompassed by an overlying family of estimators based on three-point functions of CMB temperature and galaxy clustering data.

\begin{figure}[tbh]
\centering
 \includegraphics[width=\textwidth, clip=true]{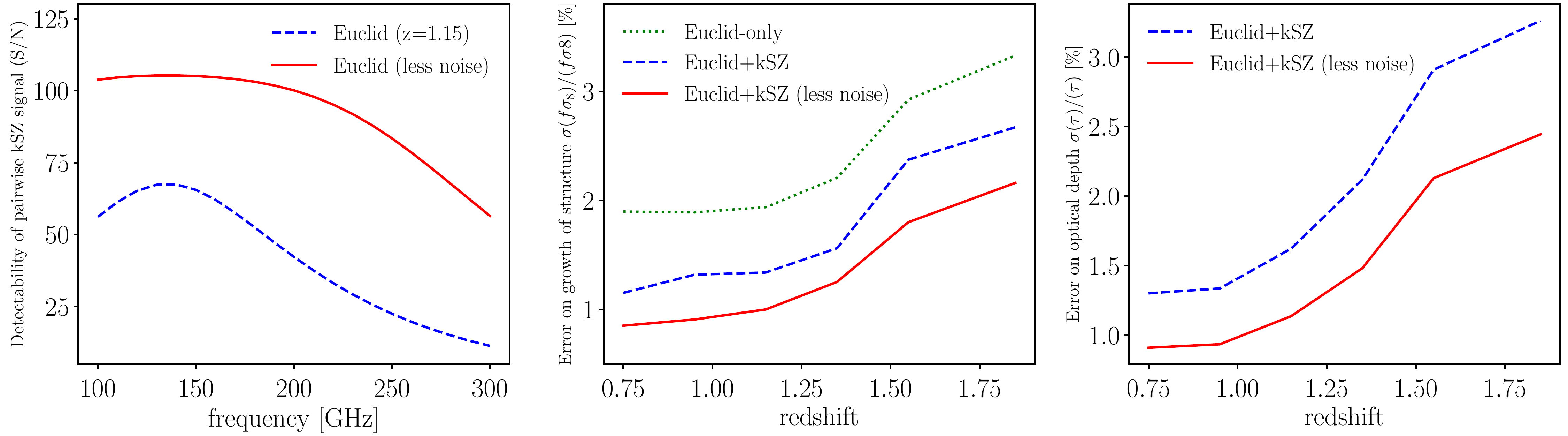}
\caption{\small 
{\it Left:} expected signal-to-noise ratios of the pairwise kSZ signal as a function of frequency. 
{\it Middle:} expected errors on the growth rate as a function of redshift.
{\it Right:} expected errors on the optical depth as a function of redshift.}
\label{fig:kSZ_forecast}
\end{figure}

Table\,\ref{table_forecasts} presents forecasts for detections of the kSZ pairwise peculiar momentum from clusters in future CMB experiments and surveys of the large-scale structure. These figures can be seen as lower bounds for the sensitivity of an ESA \mission. At large scales (i.e., in the linear regime), the pairwise velocity can be expressed as $v_{12}(r) = 2 \bar{b} \xi_{v\delta}(r)$ \citep{Keisler:2012eg}, where $\bar{b}$ is the mass-averaged halo bias (which can be measured via the cluster's auto-correlation), and $\xi_{v\delta}(r)$ is the matter-velocity correlation function. Thus, the large-scale pairwise kSZ measurement probes
$p_{\rm e}\propto \bar{\tau}_{\mathrm{eff}} (f\sigma_8)(b\sigma_8)$, where $\sigma_8$ denotes the normalization of the matter power spectrum and $\bar{\tau}_{\rm eff}$ is the effective optical depth of the sample. The detection levels shown in Table~\ref{table_forecasts} can therefore be translated into constraints on $f\sigma_8$. Note that the pairwise kSZ measurement based on galaxy surveys is affected by redshift-space distortions, which lead to small suppression of the signal at 20--100$\,$\Mpc \ and a sign inversion at $\lesssim20 \, \Mpc$ \citep{Okumura2014,Sugiyama2016,zubeldia19}.

In Fig.~\ref{fig:kSZ_forecast} we show the added value of kSZ observations when constraining the amplitude of peculiar velocities (through the parameter $f\sigma_8$) in a {\it Euclid}-type survey \citep{Euclid2011}. The left panel shows the S/N of the kSZ detection versus frequency after accounting for the noise induced by the tSZ, IR and radio contamination at each wavelength. The middle panel shows a factor of 2 improvement on the constraints on $f\sigma_8$ after adding kSZ observations, while the left panel provides the accuracy on the amount of electrons hosted by the halos in terms of the effective Thomson optical depth $\tau$. These forecasts use as a reference the H$\alpha$ emitters in the redshift range $z=0.9-2.0$ constituting Euclid's spectroscopic sample, and thus correspond to relatively low-mass halos ($\sim 10^{12}\,{\rm M}_{\odot}$).

The fact that the cosmological constraining power of kSZ is fundamentally limited by our understanding of the optical depth of galaxy groups and clusters \citep{Sugiyama:2016rue} is both a curse and a blessing. This sensitivity to $\tau$ can be used in combination with kSZ, tSZ, weak lensing, and X-ray data to constrain the impact of baryonic physics on other cosmological observables. We discuss these further in Sects.~\ref{sec:filaments} and \ref{sec:feedback}. The optical depth depends on the properties of the halo hosting the galaxy cluster, such as its mass and concentration, as well as astrophysical effects such as star-formation and feedback from active galactic nuclei (AGN) and supernovae (SNe) \citep{Flender2016,Battaglia:2016xbi,Flender2017,Soergel2018}. 
%In the absence of any other constraints, the gasphysics can introduce the variation in the optical depth (and thus the kSZ amplitude) of galaxy clusters by a factor of $\sim2$ between models with and without star-formation and feedback, the uncertainty is $\sim$100\% \citep{Flender2016}.
It is to be noted that $\tau$ could be constrained by making use of tSZ measurements, through the scaling relation between $\tau$ and $y$ \citep{Battaglia:2016xbi,Soergel2018}, as well as through
X-ray measurements of groups and clusters \citep{Flender2017} or fast radio bursts \citep{Madhavacheril2019}.
However, observational constraints on the low-mass, high-redshift objects (e.g., around galaxy groups and galaxies) are completely missing. Given the high significance of kSZ measurements expected with future experiments for these objects, a better understanding of the optical depth is thus crucial for realizing the statistical power of the upcoming galaxy and CMB surveys for cosmology, and this will be a realistic goal for the sensitive tSZ and lensing measurements outlined here for \Backlight.

%--------------
\begin{figure}[tbh!]
\centering
 \includegraphics[width=0.66\textwidth, clip=true, trim=0 270 0 270]{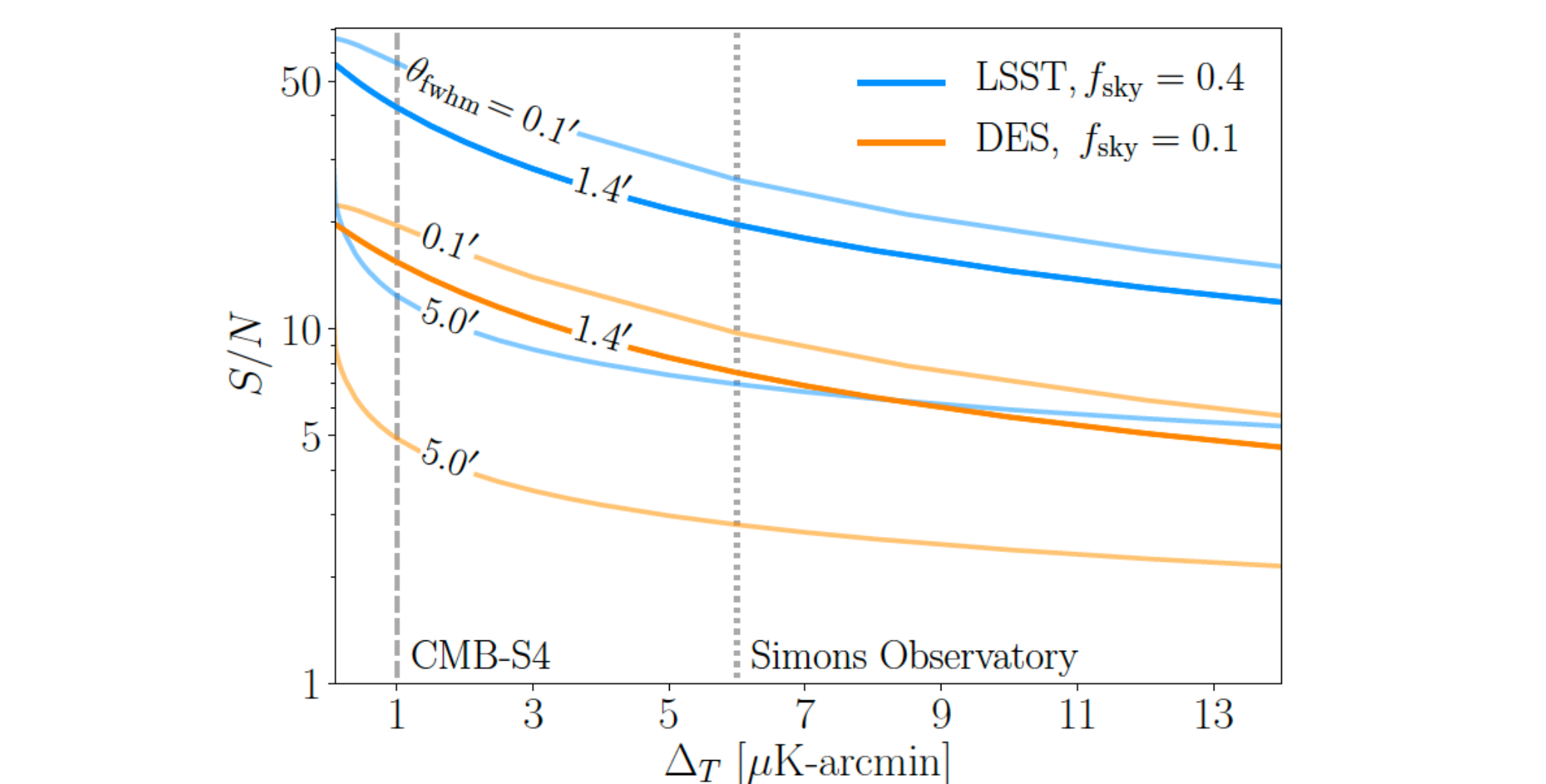}
 \caption{\small Signal-to-noise ratio for a transverse velocity estimator -- measured via the moving lens effect -- for a range of CMB noise levels and beam sizes, and combined with the DES and LSST galaxy survey data. Figure taken from Ref.~\cite{Hotinli18}. The signal-to-noise improvement due to a change in angular resolution from 5$^\prime$ to 1.4$^\prime$ is evident, whereas further decreasing the beam size produces only marginal improvement.}
\label{fig:mov-lens}
\end{figure}
%--------------

Gravitational lensing by galaxy groups and clusters can also be utilized to infer the cosmic peculiar velocity field, this time from the transverse motion via so-called ``moving lens effect" \cite{Birkinshaw1983,Birkinshaw1989,Gibilisco1997,Aghanim1998}. However, this signal is proportional to the cluster deflection angle instead of its optical depth, and due to the smallness of the former quantity, it is roughly 10--100 times smaller than the kSZ signal. In spite of that, there are realistic predictions for the detection of the moving lens effect, by constructing an optimized filter for its unique dipole-like pattern, and cross-correlating it with large galaxy catalogs expected from future optical surveys like the LSST \cite{Hotinli18,Yasini19}. In particular it has been shown \cite{Hotinli18} that for a future CMB experiment with sensitivity of $\sim 1\,\mu$K-arcmin or better, an improvement of angular resolution from 5$^\prime$ to 1.4$^\prime$ will cause an increase in the signal-to-noise ratio by roughly a factor of 4--5 (Fig.~\ref{fig:mov-lens}). On the other hand, improving the angular resolution further, to 0.1$^\prime$, will produce only marginal improvements to the signal-to-noise.

The specifications of the ESA \mission, in particular its sensitivity, angular resolution and frequency coverage, will allow us to have an unprecedented level of control over the astrophysical systematics that will ultimately limit the constraining power of both kSZ and moving-lens measurements. Contamination from the primary CMB, tSZ, radio point sources, and CIB would be mitigated to unprecedented levels, significantly improving the final sensitivity. Measurements of the kSZ and moving lens effect from \Backlight, in combination with upcoming surveys of the large scale structure, would provide the most detailed description of the 3D distribution of ionized gas in the late ($z<5$) Universe and its velocity field, and could be used to place competitive and stringent constraints on alternative dark-energy models.

\subsubsection{Search for primordial non-Gaussianity with the kSZ effect}  
\label{sec:fnl}

The primordial fluctuations that seeded cosmic structure are observed to be very close to Gaussian. Models of multi-field inflation predict a small amount of local primordial non-Gaussianity, parameterized by the amplitude $f_\mathrm{NL}$. In general, all single-field inflation models with standard Bunch-Davies initial conditions can be ruled out by a detection of  $|f_\mathrm{NL}|\gtrsim 1$ \citep{Maldacena2003,Creminelli2004}\
. This makes searches for non-Gaussianity a powerful probe of the early Universe. 

In addition to inducing higher-point functions in late time observables, this type of local primordial non-Gaussianity leads to a distinct scale-dependence of galaxy bias on large scales, ranging as a function of comoving wavenumber $k$ as $f_\mathrm{NL}/k^2$ \citep{0710.4560}. Non-Gaussianity can thus be detected or constrained by measuring galaxy clustering on large scales. Such a measurement is, however, limited by sample variance in the fluctuations on large scales. If a different measurement that traces the same underlying matter density field is available, a cross-correlation analysis would allow us to cancel part of the sample variance and improve the constraint. Joint analyses of CMB lensing and galaxy clustering data have been suggested as a route towards such improved constraints \cite{Schmittfull2018}, although the breadth in redshift of the CMB lensing kernel reduces the cross-correlation coefficient and limits possible improvements. 

KSZ velocity reconstruction can be used as an alternative measurement of the unbiased large-scale density fluctuations. The kSZ effect allows reconstruction of the radial velocity field from a galaxy survey and CMB temperature measurements. This reconstruction can be made up to an overall scale-independent amplitude that depends on small-scale astrophysics \citep{Smith2018}. On large scales where linear theory is valid, velocities directly trace the underlying matter density field, thus providing an alternate view of the density modes. Since the kSZ velocity reconstruction has an uncertain normalization, direct inference of the growth rate or the amplitude of structure formation is limited by prior knowledge of the small-scale astrophysics (often called the ``optical depth of galaxies''). However, scale-dependent effects such as the impact of primordial non-Gaussianity on the galaxy bias can be measured. In this way, cross-correlating the kSZ velocity reconstruction and the galaxy density improves the $f_{\mathrm{NL}}$ constraint when the galaxy clustering measurement is sample variance limited, since the scale-dependence of the galaxy bias can be directly measured with part of the sample variance canceled. As shown in Ref.~\citep{Munchmeyer2018}, factors of a few improvement over the galaxy survey alone is possible when combining LSST with Simons Observatory ($\sigma(f_{\rm{NL}})\approx 1$) or CMB-S4 ($\sigma(f_{\rm{NL}}) \approx 0.5$). Further significant improvements can be made when combining a full-sky survey like SPHEREx \cite{Spherex} and \Backlight, both because of the larger number of modes available from the full-sky coverage and because of the higher sensitivity in the CMB temperature map thanks to the increased frequency coverage, reducing the foreground residuals in the map.

\subsection{The Cosmic Web: Relation Between Dark Matter and Baryons}

\subsubsection{SZ polarisation (pSZ)}
\label{sec:pSZ}

While Compton-scattered photons are linearly polarized,
a quadrupole component in the CMB angular anisotropy (as seen by electrons in the cluster) is needed in order to produce any net polarization in the SZ signal \citep{SZ1980,ZS1980}. Therefore, SZ polarization provides an important diagnostic on the local CMB anisotropy specific to the studied cluster. Such anisotropy can be inherent to the primary CMB radiation field at cluster locations \citep{SZ1980,1997PhRvD..56.4511K,SS1999}, as well as be induced by the cluster itself, for instance by its bulk motion with respect to the CMB or by the second-scattering effects \citep{ZS1980,SS1999,2003ApJ...597L...1D,Lavaux2004,Shimon2006}. Also, the local CMB intensity distribution can be distorted by gravitational effects, e.g., the moving-gravitational-lens effect  discussed above (Sect.~\ref{sec:kSZveloc}), or rotation of the cluster~\citep{Chluba2002}.

%\clearpage
%==============================================================================
\begin{figure}[tbh]
\centering
\includegraphics[scale=0.5, clip=true, trim=0 175 0 100]{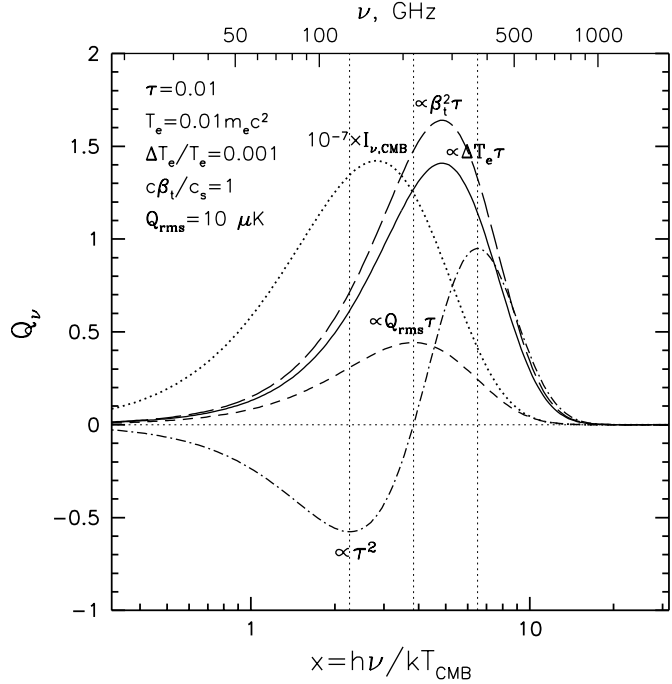}
\caption{Relative amplitudes and spectral dependencies of various polarization signals compared to the CMB intensity multiplied by $ 10^{-7}$ (dotted curve) for a cloud of electrons with temperature $ T_{\rm e}=0.01m_{\rm e}c^2=5.1$\,keV and Thomson optical depth $\tau=0.01$. The effects due to electron pressure anisotropy at the level $ \Delta T_{\rm e}/T_{\rm e}=10^{-3}$ and bulk transverse motion with velocity $ \beta_{\rm t}c$ (with respect to the CMB radiation field) equal to the adiabatic speed of sound in the cloud ($c_{\rm s}=\sqrt{\gamma T_{\rm e}/\mu m_{\rm p}} $, $\gamma=5/3$ and $\mu=0.6$) are shown as solid and long-dashed curves, respectively. The effects due to second scatterings and the intrinsic CMB quadrupole (with rms amplitude $ Q_{rms}=10 \mu K$) are shown as dash-dotted and short-dashed curves. The vertical dotted lines mark $x=h\nu/kT_{\rm CMB}=2.26$ (128\,GHz), 3.83 (218\,GHz), and 6.51 (370\,GHz). Figure adapted from \cite{2018MNRAS.474.2389K}.
}
\label{f:spec}
\end{figure}
%==============================================================================
%==============================================================================
% The table
%------------------------------------------------------------------------------------------
%=============================================================================
\begin{table*}[tbh]
\caption{Summary of various sources of CMB polarization in the direction of galaxy clusters. The polarization degree is expressed as $P=P_0\alpha(T,\tau,\beta_t,...)\varphi(x)$, where $ P_{0}$ is the amplitude calculated for a fiducial set of parameters (see below), $\alpha(\tau,\beta_t,...)$ the scaling of the amplitude with these parameters, and  $\varphi(x)$, where $x=h\nu/kT_{\rm CMB}$, describes the spectral dependence of the signal. The fiducial set of parameters is $\tau=0.01$, $ kT_{\rm e}=0.01m_{\rm e}c^2=5.1$\,keV, $ \Delta T_{\rm e}/T_{\rm e}=10^{-3}$, $ Q_{\rm rms}=10~\mu K$,  $ \beta_{\rm t}c=1000\,{\rm km}\,{\rm s}^{-1}$, $\beta_{\rm r}c=100\,{\rm km}\,{\rm s}^{-1}$, $\Delta \theta=1$ arcmin, and $D^{EE}_\ell=0.1\,\mu$K at $\ell=10^4$. Here, $Q_{\rm rms}$ is the rms amplitude of the local CMB quadrupole component (e.g., \cite{Bennett2003}), $ \beta_{\rm t}\,c$ the cluster's transverse bulk velocity, $\beta_{\rm r}c$ the circular velocity due to rotation of the cluster \citep{Chluba2002}, $\Delta\theta=4G M_{\rm cl}/c^2R\approx0.7$ arcmin $ (M_{cl}/10^{15}{\rm M}_{\odot})(1\,{\rm Mpc}/R)$  the angle of gravitational deflection of CMB photons by a cluster of mass $ M_{\rm cl}$ at impact parameter $R$ \citep{Gibilisco1997}, $D^{EE}_\ell=\ell(\ell+1)C^{EE}_\ell/2\pi$ is the $E$-polarization power spectrum amplitude at $\ell\approx10^4$ \citep{Lewis2006}. Table adapted from \cite{2018MNRAS.474.2389K}.}
%\begin{tabular}{ccccc}
\smallskip
\centering
\begin{tabular}{lccll}
\hline\hline
Effect causing & Fiducial  & {Scaling} & Spectral  & \multirow{2}{*}{Reference}\\
polarization &   level $P_0$ & $\alpha(\tau,\beta_t,...)$  & shape $\varphi(x)$  & \\
\hline
\smallskip
CMB quadrupole & $10^{-8}$& $\propto \frac{Q_{\rm rms}}{T_{\rm CMB}}~\tau$  & $\frac{x e^x}{e^x-1}$ & \cite{SZ1980}\\
\smallskip
Bulk transverse motion &  $10^{-8}$ &$\propto \beta_{\rm t}^2~\tau$ & $\frac{e^{x}(e^{x}+1)}{2(e^{x}-1)^2}x^2$ &\cite{SZ1980}\\
\smallskip
Second scatterings ($\tau^2$) & $10^{-8}$& $\propto \frac{kT_{\rm e}}{m_{\rm e} c^2}~\tau^2$  & $\frac{x e^{x}}{e^{x}-1}\left(x\frac{e^{x}+1}{e^{x}-1}-4\right)$ &\cite{SZ1980}\\
\smallskip\smallskip
\smallskip
Bulk transverse anisotropy &  $10^{-8}$ &$\propto \left<\beta_{\rm t}^2\right>~\tau$ & $\frac{e^{x}(e^{x}+1)}{2(e^{x}-1)^2}x^2$ &\cite{2003ApJ...597L...1D}\\
\smallskip
Pressure anisotropy & $10^{-8}$ & $\propto \frac{\Delta T_{\rm e}}{T_{\rm e}}\frac{kT_{\rm e}}{m_{\rm e} c^2}~\tau$  & $\frac{e^{x}(e^{x}+1)}{2(e^{x}-1)^2}x^2 $ & \cite{2018MNRAS.474.2389K}\\
\smallskip
Moving lens & $10^{-9}$& $\propto \beta_{\rm t}\Delta\theta~\tau $  & $\frac{x e^x}{e^x-1}$&\cite{Birkinshaw1983}\\
\smallskip
Cluster rotation & $10^{-10}$& $\propto \beta_{\rm r}^2 ~\tau $  & $\frac{e^{x}(e^{x}+1)}{2(e^{x}-1)^2}x^2$& \cite{Chluba2002}\\
%CMB fluctuations & $\propto \sqrt{l(l+1)C_{EE}(l)}|_{l~3000}$ & $\sim 10^{-8}$ & $\varphi_{0}(x)$ \\
\smallskip
CMB fluctuations & $10^{-8}$& $\propto \frac{\sqrt{D^{EE}_\ell}}{T_{\rm CMB}}$  & $\frac{x e^x}{e^x-1}$ &\cite{Lewis2006}\\
%CMB lensing & $\propto \mu\sqrt{D^{EE}_{l}}$ & $\sim 10^{-8}$ & $\varphi_{0}(x)$ \\
\hline
\end{tabular}
\label{t:effects}
\end{table*}
%=============================================================================

The amplitude of the bulk-motion-induced anisotropy is typically a factor of $m_{\rm p}/m_{\rm e}\approx2000$ smaller than the amplitude of the thermal SZ effect from the same cluster, making its detection very challenging. Actually, what is important is not the anisotropy of the bulk motion itself, but the anisotropy of its velocity squared, which might be present even in case of zero total bulk motion, e.g., for a two-clusters merging perpendicular to the line of sight. Such a situation might also take place if the distribution function of the thermal electrons is not fully isotropic, e.g., {bi-Maxwellian}. The ICM plasma is weakly-collisional and magnetized, so a certain degree of electron pressure anisotropy might be induced by magnetic field stretching and thermal conduction \citep[e.g.,][]{Schekochihin2006}. In merging clusters, such anisotropies can be generated in a correlated manner across large volumes of the ICM and give rise to a specific SZ polarization signal \citep{2018MNRAS.474.2389K}. 

{Remarkably, the amplitudes of the polarization signals induced by the different mechanisms mentioned above} appear to be comparable to each other, amounting to $\sim10^{-8}$ relative to the primary CMB radiation field (see Table \ref{t:effects} and Fig.~\ref{f:spec}).  In order to detect the SZ polarization on top of various contaminants and disentangle the complex mixture of the effects contributing to it, very sensitive (down to about the 10-nK level) CMB polarization measurements need to be obtained in several frequency bands, which requires space-based missions. Indeed, the individual polarization effects have either distinct morphologies 
%(e.g., the bulk motion SZ polarization comes mainly from the gas of the denser and colder sub-structures, inside which e.g. the anisotropy-induced polarization is negligible due to higher collisions frequency in such gas), 
or distinct spectral shapes 
%(e.g., the $\tau^2$ polarization has similar spectral dependence as the thermal SZ effect, so it should vanish around $x=h\nu/kT_{cmb}=3.83$ (218 GHz), 
(see Fig.~\ref{f:spec}). Requirements for the angular resolution of these data are relatively mild, since the regions of the correlated polarization signals are predicted to be quite extended, comparable to the size of the cluster itself. %(but potentially with very sharp boundaries). 
Synergy with the next-generation X-ray observatories looks especially promising in this context -- X-ray data are capable of providing hydro- and thermo-dynamical properties of the ICM plasma flows, while SZ polarisation will probe such intricate physical processes like plasma kinetics, thermal conduction, turbulence, and magnetic fields evolution in the ICM. The new facilities will open a unique possibility to simultaneously study phenomena separated by some 20 orders of magnitude {in spatial} scale, i.e., ranging from the CMB quadrupole anisotropy (and its variation across the sky and in redshift) down to the ICM plasma kinetics {(operating at scales comparable to the particles' Larmor radii).}

\subsubsection{Baryons in cosmic filaments with tSZ and kSZ} 
\label{sec:filaments}

\noindent The cycling of baryons in the Universe is intimately related to galaxy formation and evolution. Yet, despite the impression that detecting baryons would be a much easier task than exploring the dark sector of the Universe, we have not achieved a complete observational census of all the distinct phases of the baryonic matter. Perhaps 40\% of the baryons predicted by Big Bang Nucleosynthesis, and subsequently measured at the epoch of recombination through the CMB temperature power spectrum or at high redshifts ($z>2$) via the Ly-$\alpha$ forests, is unaccounted for in the observations of the local Universe. This is known as the ``missing baryon problem'' \cite{Cen1999,Shull2012,Nicastro2018}. Most of these unobserved baryons lie outside the boundaries of collapsed, virialized halos \cite{Planelles2018,Martizzi2019} and the tSZ and kSZ effects are new probes for these diffuse baryonic phases with significant future promise. 

%%%%%%%%
\begin{figure}[bthp!]
\centering
%\begin{center}
  \includegraphics[width=\textwidth, clip=true, trim=0 160 0 160]{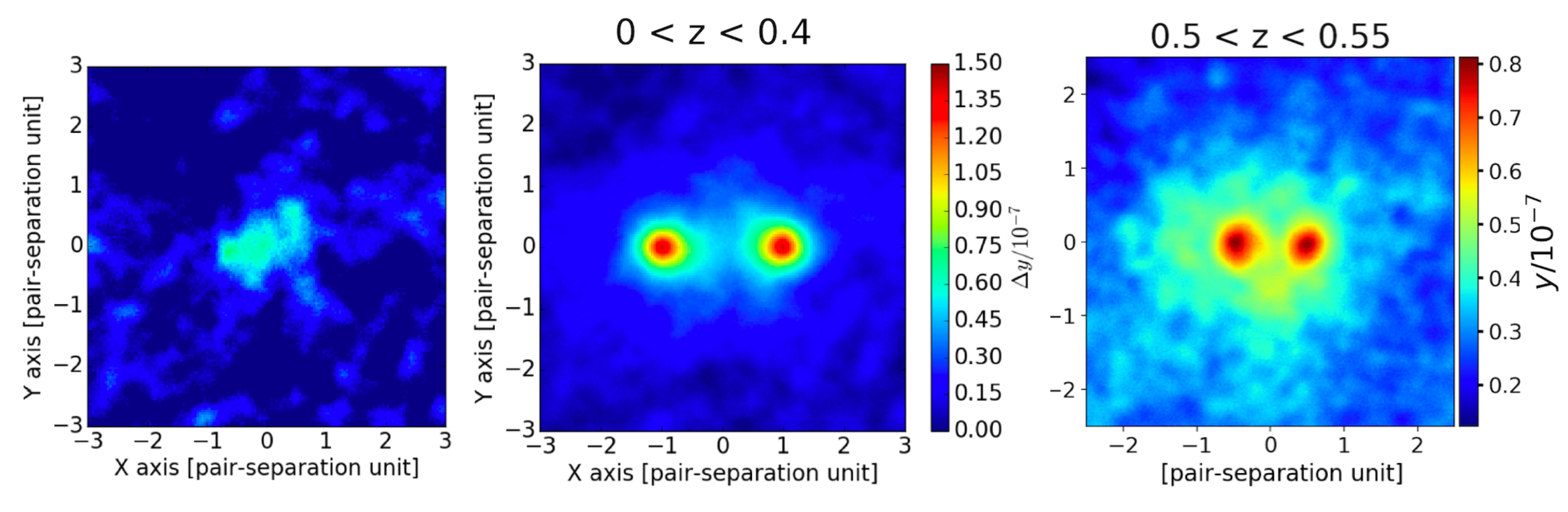}
%\end{center}
\caption{\small  Recent measurements of the warm-hot baryonic phase in cosmic filaments via the tSZ effect. The two left panels are from Ref.~\cite{Tanimura2019} who stacked roughly 260,000 galaxy pairs to obtain a $5\sigma$ detection of the tSZ excess coming from the unbound gas in-between (left panel). This result is based on {\it Planck\/} sky maps, with barely enough resolution for separating the halo and filament contributions at low redshifts (middle panel), whereas at higher redshifts (e.g., at $z\sim 0.5$, right panel) the halo pairs become partially unresolved, which prohibits the study the redshift evolution of this warm-hot phase in further detail.}
\label{fig:filaments}
\end{figure}
%%%%%%%%%

A large fraction of the unobserved baryons is believed to exist in a warm-hot phase ($10^5\;\mathrm{K} <  T_{\rm e} <10^7\;\mathrm{K}$) along cosmic filaments and is traditionally investigated via quasar absorption lines in the UV and X-rays. Recent works have shown the promise of the tSZ and kSZ effects for its detection \cite{prl_chm15,prl_hill16,Tanimura2019,DeGraaff2019}. However, the detection significance remains low and any major advances in our understanding of the physical properties and time evolution of this phase will need an order of magnitude improvement in sensitivity and resolution (Fig.~\ref{fig:filaments}). Another extremely important phase of the missing baryons is the cold circumgalactic medium (CGM; $T_{\rm e} \lesssim 10^5\;\mathrm{K}$) which forms the interface between the dense interstellar medium and the cosmic web, playing a crucial role in the formation and evolution of galaxies. Stacking analyses with the {\it Planck}, SPT, and ACT data have shown the promise of the tSZ effect to study the thermodynamic state of the ionized gas around halos down to galaxy mass scales \cite{PlanckXI2013,Greco2015,2015JCAP...09..046M}. Improved sensitivity and large sky coverage of future space-based experiments will probe the physical state of the circumgalactic medium in a wide redshift range by means of stacking and cross-correlation analyses, advancing our knowledge of galaxy formation and feedback processes (\cite{Battaglia2017,Battaglia_Astro2020}, see \ref{sec:feedback}). One critical advantage of a space mission will be its dense frequency coverage up to $\sim 1$\,THz, which can effectively separate the $y$-maps from contamination by the thermal dust emission of galaxies. As tSZ observations from both ground and space will cross new sensitivity thresholds in the future, separation of the astrophysical contaminants will be \emph{the} determining factor in making scientific progress.

An even more promising avenue for studying the cosmic baryons in low-density regions is through the kSZ effect. Since the kSZ signal is proportional to the total column density of free electrons, irrespective of their temperature, it is not biased towards measuring only the hot gas in the Universe, as with the tSZ effect. The kSZ is thus sensitive to {\it all} baryons that lie within the same co-moving region (of $\sim 40$--60\,Mpc typical size) feeling a similar gravitational attraction. This has prompted several studies focusing on the kSZ effect to address the missing baryon problem  \citep{DeDeo:2005yr,Hernandez08,Ferraro2016,chm_ho_09,ho_09,Ade:2015lza,Park_2018}, and actually claiming the detection of the missing baryons around local galaxies \citep{prl_chm15} and even up to $z\simeq 2$ \citep{prl_hill16}. 
%%%%%%%%%
Here the challenge lies in separating the kSZ signal from the primordial CMB fluctuations, which shares the same frequency spectrum (to the first order) but dominates over the kSZ power at $\ell< 4\,000$ (see, e.g., \citep{chm_ho_09}). The angular resolution of a CMB experiment is key in this context to separate kSZ anisotropies from those generated at $z\simeq 1100$; as shown in figure~6 of Ref.~\citep{HernandezMonteagudo:2005ys}, the CMB contamination to kSZ measurements drops dramatically for angular resolutions of 3--4\,arcmin (and finer). This requirement would be fulfilled by \Backlight\ (FWHM$\sim 1$--$1.5\,$arcmin), which, combined with its multi-frequency information, would enable an exquisite mapping of the kSZ and tSZ effects, shedding unprecedented light on the spatial distribution and the thermo-dynamical state of the gas at a wide range of recent cosmological epochs ($z<5$). 

This type of study has only been partially conducted in the (very) nearby Universe for which \Planck's angular resolution at 5\,arcmin suffices. The extension of these studies up to higher redshifts would provide insights into the process of gas flows along filaments and its accretion onto halos, and on the impact of super-massive black hole, SNe, and galactic winds in the build-up of halos in different mass ranges. This is a long-standing problem lying at the heart of our understanding of galaxy formation and evolution, while having cosmological implications related to the presence of light particles and the growth of structure at late epochs. 
%%%%%%%%
Once again, the defining advantage of a future space mission will be its comprehensive frequency coverage and accurate channel calibration, to facilitate precise separation of astrophysical components and the construction of CMB maps (containing the kSZ signal) that are free from other tracers of large-scale structures (such as dust emission and tSZ). 

\subsubsection{Feedback modeling from joint tSZ+kSZ+lensing} 
%\textcolor{red}{[J. C. Hill, N. Battaglia]}
\label{sec:feedback}

%In the coming decades, the major goal for galaxy formation theory is understanding the physical processes within and thermodynamic properties of the ionized baryons in galaxies and clusters. 
In the coming decades, a major goal for galaxy formation theory will be to gain understanding of the physical processes that determine the  thermodynamic properties of the ionized baryons in galaxies and clusters. Hardly any of the baryons in these objects are located in stars, but rather in the circumgalactic medium (CGM) and the intracluster medium (ICM).  The CGM and ICM are the baryonic reservoirs that dictate the evolution of galaxies and clusters.  Encoded in their thermodynamic properties are the effects of feedback processes that shape galaxy and cluster formation.  A precise understanding of the energetics of feedback in the CGM and ICM is required to constrain the space of galaxy formation models~\cite{vandeVoort2016,Battaglia2017,Battaglia_Astro2020}.

Several feedback processes must be understood in the CGM and ICM.  In the standard picture, halos with masses near to or below that of the Milky Way are primarily affected by feedback from stellar winds and supernovae (``stellar feedback'')~\citep[e.g.,][]{RS2015}.  In more massive halos, including groups and clusters, feedback from AGN is expected to be the primary mechanism regulating star formation, although stellar feedback could have played an important role at earlier times.  At present, current implementations of these processes in large-scale simulations can reproduce the observed stellar properties of galaxies, but predictions for the properties of ionized gas vary widely~\citep[e.g.,][]{fire,Eagles,HorizonAGN,tng2018}.  In particular, given the wide range of scales involved, it is necessary to probe the CGM and ICM over broad windows in halo mass, redshift, and spatial separation in order to make breakthroughs in this area (see Fig.~2 in \cite{Battaglia_Astro2020}).

Measurements of the tSZ, kSZ, and CMB lensing signals using \Backlight\ will have the capability to provide these breakthroughs, not only on their own, but also in tandem with expected arcmin-resolution measurements at $\lesssim 200$\,GHz from the ground using CMB-S4 and other facilities.  As shown earlier, \Backlight\ would detect all halos with mass greater than $5 \times 10^{13} {\rm M}_{\odot}$ in the entire Universe; in addition, with its full-sky coverage, it would overlap with all existing halo samples at other wavelengths, including \Euclid\, LSST, \eROSITA\, and others.  Thus, tSZ, kSZ, and lensing measurements would be possible over a vast range in halo mass and redshift.  These data would provide a complete census of the thermodynamic properties of the ionized gas in these objects, including the pressure, temperature, and density profiles (gas and dark matter).  As shown in Ref.~\cite{Battaglia2017,Battaglia_Astro2020}, such measurements would provide crucial constraints on our understanding of feedback mechanisms in galaxy formation and evolution, likely ruling out all current models for these phenomena, and pointing the way toward a deeper understanding of the relevant processes.  While the angular resolution of \Backlight\ would be limited for low-mass halos ($\lesssim 10^{13} {\rm M}_{\odot}$) at $z>1$, its high-frequency arcmin-resolution data would provide crucial complementarity with low-frequency arcmin-resolution data from CMB-S4, allowing the clean separation of multi-component tSZ, kSZ, dust, and radio source signals.  At high redshifts, the tSZ and kSZ signals are likely to be our only hope to probe the ionized gas in galaxy-scale halos (e.g., due to surface brightness dimming in X-rays); \Backlight\ would thus be an essential tool in resolving some of the most pressing questions in galaxy formation physics.

%------------------------------------------
\subsection{State of the Baryons: Relativistic Plasmas and Early Metals}
%\vspace{-3mm}
\subsubsection{Relativistic SZ effect (rSZ)} 
\label{sec:rSZ}
%\vspace{-1mm}
%------------------------------------------
While the ``first SZ revolution" happened with the third-generation CMB space mission \textit{Planck}, through the mapping of the tSZ Compton-$y$ parameter across the entire sky \cite{Planck2015XXII}, and the inference of cluster pressure profiles \cite{PlanckIntV2013}, we envision that a ``second SZ revolution"  will occur with next-generation CMB space missions through the mapping of the electron gas temperature $T_{\rm e}$ across the entire sky and the inference of cluster temperature profiles thanks to the measurement of the rSZ effect \cite{Remazeilles2019b,Basu2020Decadal}, therefore opening a new spectroscopic window on galaxy clusters not only across frequency but now also across temperature, and providing a new map-based observable $T_{\rm e}$, in addition to $y$, to constrain cosmology.

With typical masses ${M \gtrsim 4\times 10^{14}\,{\rm M}_\odot}$, galaxy clusters are the most massive bound objects in the Universe, hence they are hot, with a typical average temperature of ${kT_{\rm e} \gtrsim 5}$\,keV \cite{Refregier2000,Komatsu2002,Erler2018,Remazeilles2019} for the virialized gas of electrons residing in them. Therefore, electrons in galaxy clusters are \textit{relativistic}, with thermal velocities reaching a significant fraction of the speed of light ${v_{\rm e}^{\rm th} = \sqrt{2\,k\,T_{\rm e} / m_{\rm e}} \gtrsim 0.1\,c}$. Small relativistic temperature corrections to the tSZ effect (rSZ effect) thus come into play \cite{Chluba2013moments,Challinor1998,Itoh1998,Sazonov1998,Chluba12}, making the actual spectral signature (SED) of the tSZ effect dependent on the local electron temperature $T_{\rm e}$ (Fig.~\ref{fig:ground-space}; \textit{left}), hence varying not only across frequency but also across the sky.

Relativistic temperature corrections to the tSZ effect have still not been detected for individual clusters due to lack of sensitivity of third-generation CMB experiments. As shown by Ref.~\cite{Remazeilles2019b}, a future CMB space mission with average sensitivity $\lesssim 1\,\mu\text{K-arcmin}$ and broad spectral coverage probing high frequencies $\gtrsim 300$\,GHz, like \textit{PICO} \cite{PICO2019}, would allow us to clean the astrophysical foregrounds and disentangle for the very first time the $y$ and $T_{\rm e}$ observables of the rSZ effect across the entire sky.

\begin{figure}[tbh]
\centering
%\begin{center}
  \includegraphics[width=0.55\textwidth]{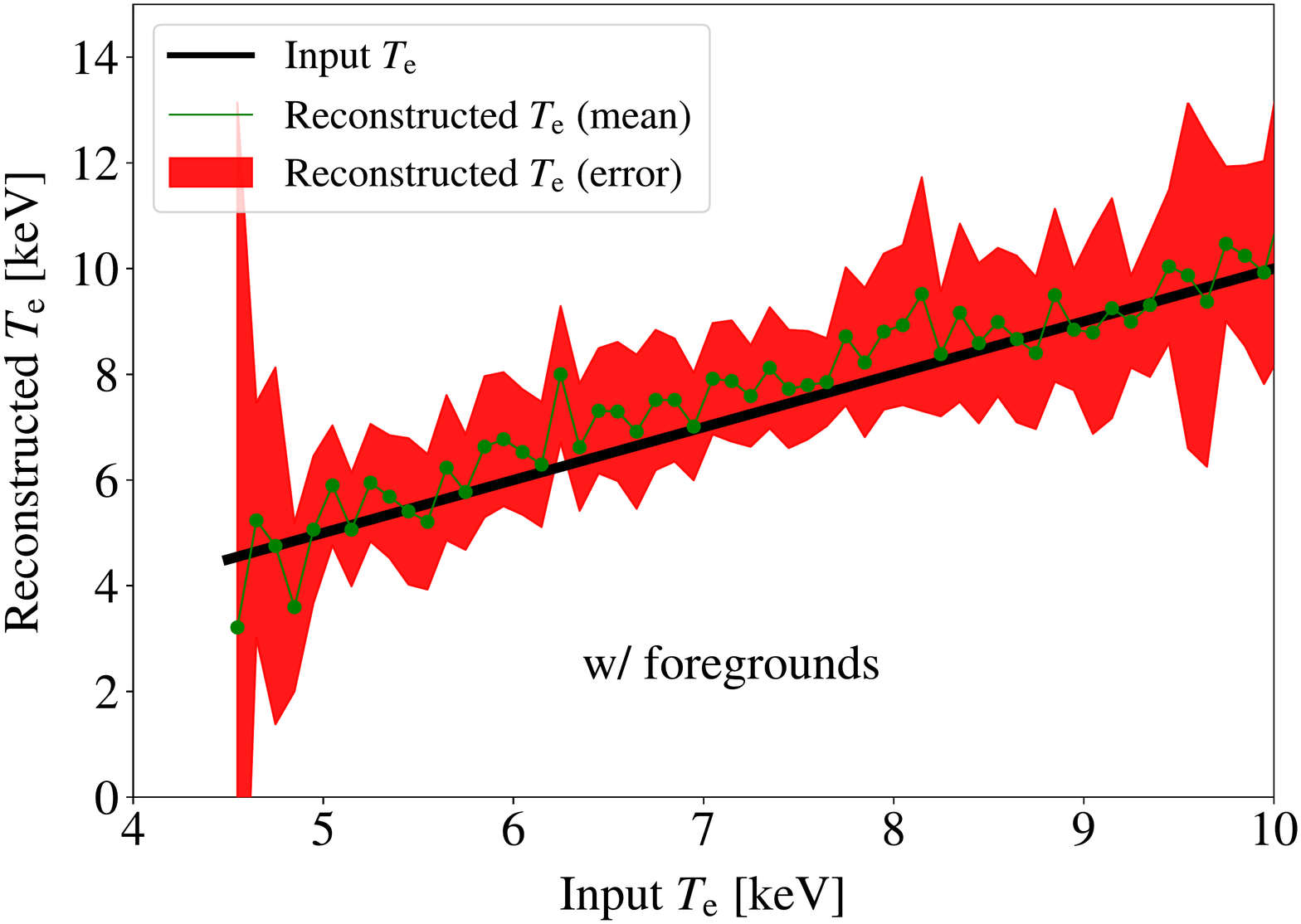}
  
  \vspace{4mm}
  
   \includegraphics[width=0.45\textwidth]{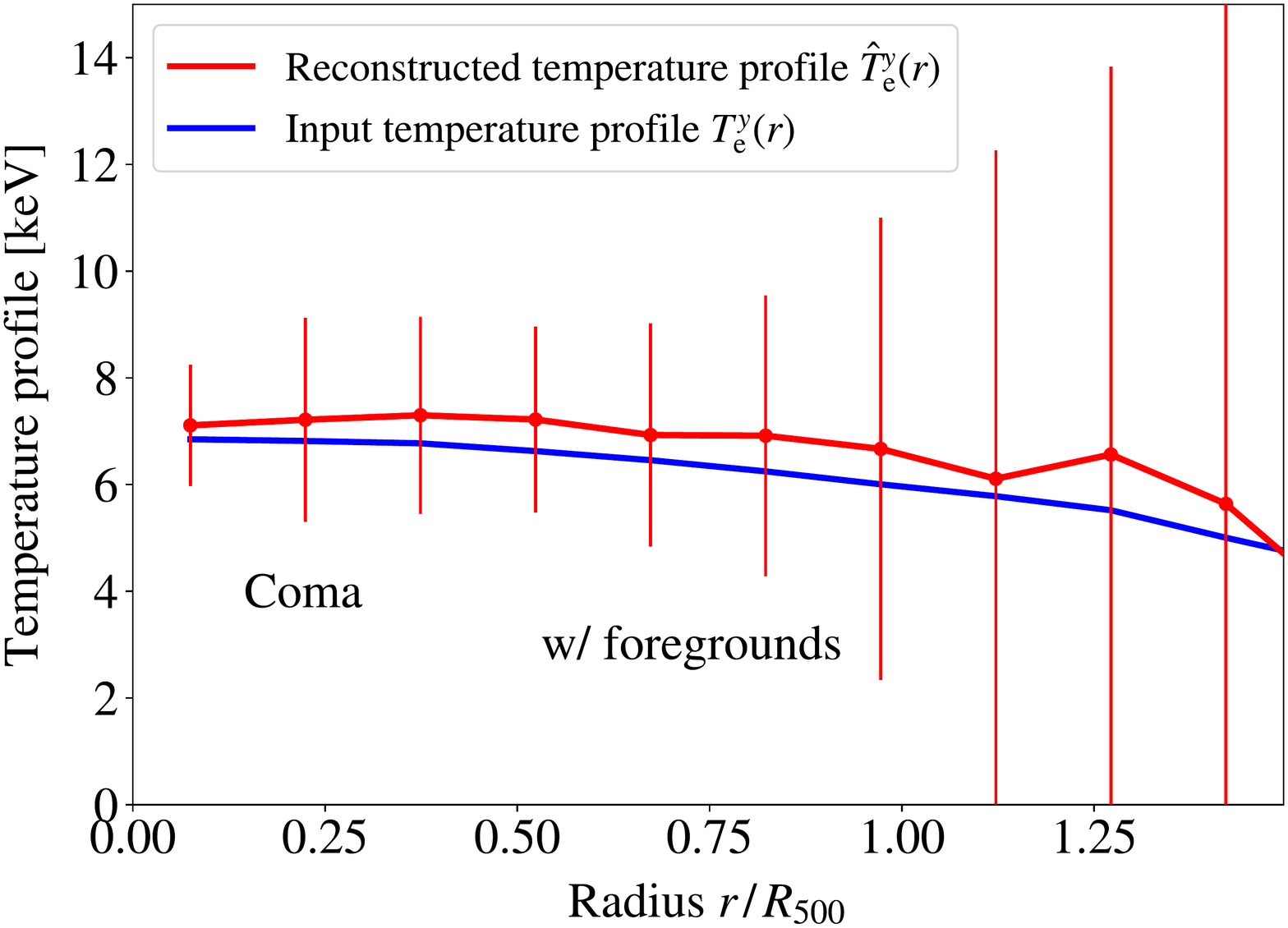}%
    \includegraphics[width=0.45\textwidth]{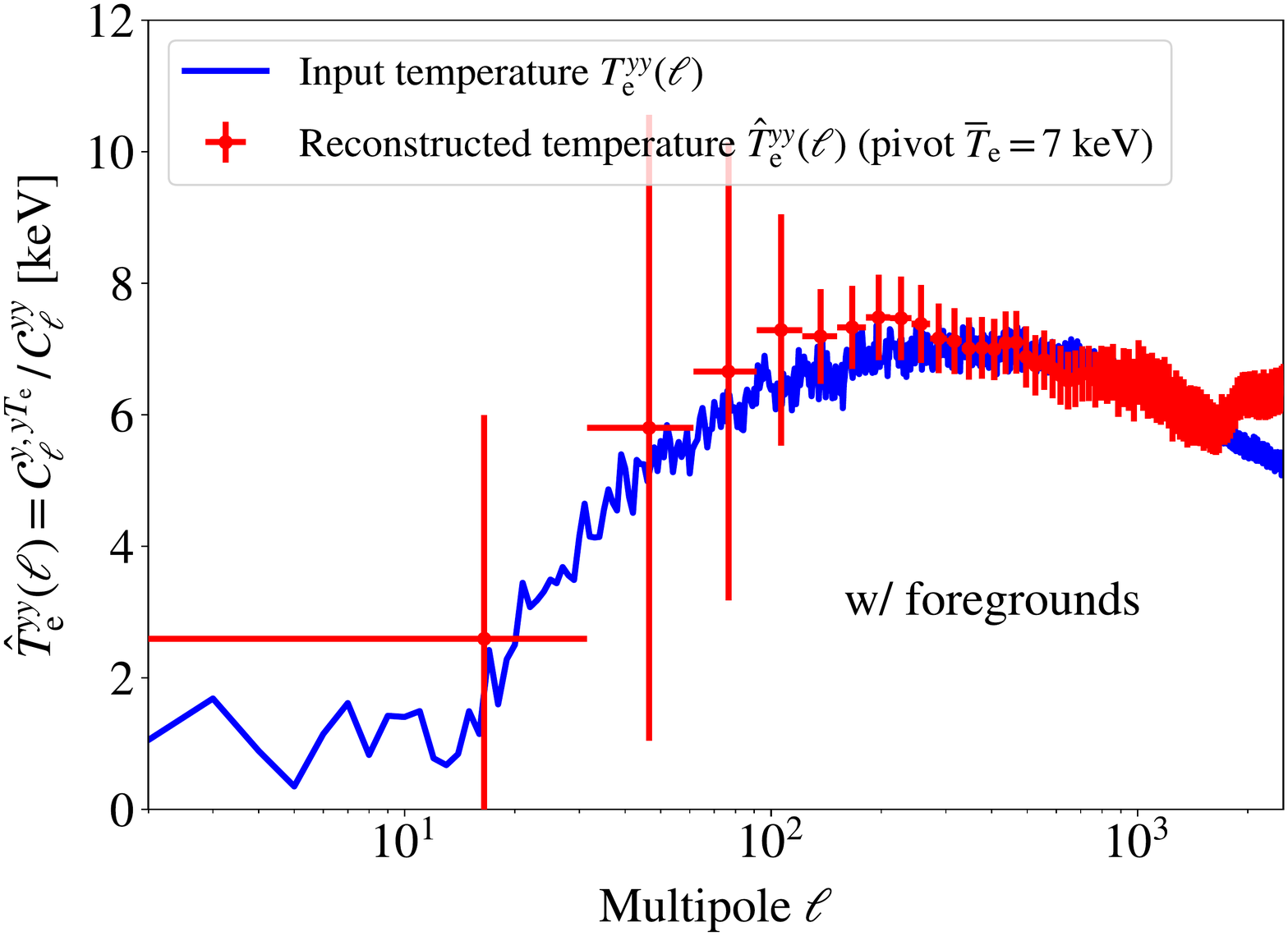}~
%\end{center}
\caption{\small {\it Top:} Reconstructed electron temperatures $T_{\rm e}$ of a large sample of clusters across the entire sky after foreground cleaning for a \textit{PICO}-type experiment. {\it Lower left:} Reconstructed temperature profile $T_{\rm e}(r)$ of the Coma cluster after foreground cleaning. {\it Lower right:} Reconstructed diffuse electron gas temperature power spectrum, $T_{\rm e}^{yy}(\ell)$, across angular scales after foreground cleaning. Figures from Ref.~\cite{Remazeilles2019b}. 
}
\label{fig:te-mapping}
\end{figure}

Figure~\ref{fig:te-mapping} shows some forecasts on the rSZ effect after foreground removal and component separation of the $y$ and $T_{\rm e}$ observables, which would be achieved with a space mission of sensitivity $\lesssim 1\,\mu$K-arcmin and $20$--$800$\,GHz spectral coverage. We detail hereafter the scientific impact on cosmology and astrophysics of mapping rSZ electron temperatures across the entire sky.

In the top panel of Fig.~\ref{fig:te-mapping}, we show the whole-sky recovery of the electron gas temperatures of a large sample of clusters after foreground removal and rSZ component separation. The recovery of electron temperatures of a large sample of clusters across the sky will offer a new independent proxy for determining the cluster masses directly from the rSZ effect, i.e., without having to rely on X-ray scaling relations which generally underestimate the actual virial temperature of the clusters \cite[e.g., Ref.][]{Pointecouteau98,Kay08}. 

In the lower left panel of Fig.~\ref{fig:te-mapping} we show the reconstructed rSZ temperature profile of the Coma cluster, which would be measured at $10\sigma$ significance after foreground removal with such spectral coverage. The measurement of individual cluster temperature profiles will break the electron density-temperature degeneracy in the cluster pressure profiles, hence probing the missing baryons in the Universe through the inferred electron density profiles. As a direct consequence, the inferred electron density profiles from the rSZ will also enable measurements of the cluster velocity profiles through the kSZ effect. This will significantly improve our understanding of the full thermodynamic properties of galaxy clusters.

In the lower right panel of Fig.~\ref{fig:te-mapping}, we show the reconstruction of the \textit{diffuse} electron gas temperature power spectrum $T^{yy}_{\rm e}(\ell)$ \cite{Remazeilles2019}, i.e., the \textit{average} electron temperature over the sky across different angular scales, after foreground removal. This will deliver a new map-based observable, the $T_{\rm e}^{\,yy}(\ell)$ power spectrum, whose shape and amplitude depend on cosmological parameters in a different way than the $y$-map power spectrum $C_\ell^{yy}$ \cite{Planck2015XXII}, hence it will complement the $y$-map power spectrum to constrain cosmology with galaxy clusters and break several cosmological parameter degeneracies, possibly alleviating some of the current tensions between primary CMB and cluster cosmology results \citep{Remazeilles2019,Remazeilles2019b}.

High-frequency coverage $\gg300$\,GHz is essential to measure rSZ temperatures without confusion, since it is the only way to overcome the temperature degeneracy in the tSZ spectrum at lower frequencies (Fig.~\ref{fig:ground-space}; \textit{left}). High-frequency coverage for rSZ studies can reliably be achieved from space, hence the strong motivation for a \mission. Besides the high-frequency coverage requirement strictly for detection, a comprehensive spectral coverage of $20$--$1000$\,GHz would ensure confident cleaning of the various foregrounds to extract the faint rSZ signal without bias (see \cite{Basu2020Decadal}). Broad spectral coverage is also essential to obtain evidence for incorrect foreground modeling and false signal detections, since it can overcome the degeneracy between different foreground models that occurs over narrower frequency ranges \cite{Remazeilles2016, Hensley2018}. The false detection of primordial $B$ modes by the ground-based experiment BICEP2 \citep{BICEP2_2014}, subsequently attributed to Galactic dust thanks to \textit{Planck}'s 353-GHz observations, serves as a reminder of the importance of space observations at high frequencies to properly characterize residual foreground contamination. In addition, a relatively high resolution ($\lesssim 1.5'$ at frequencies $\gtrsim 300$\,GHz) would resolve most rSZ clusters with masses ${M \gtrsim 10^{14}\, {\rm M}_\odot}$. Finally, the full-sky survey allowed by \Backlight\ would allow us to map the diffuse electron gas temperature at large angular scales across the sky, thus offering a new map-based observable to constrain cosmology with galaxy clusters.

\subsubsection{Non-thermal relativistic SZ for cosmic-ray energy budget} 
\label{sec:nonthermalrsz}

For the SZ effect, the scattering electrons can also have non-thermal velocity distributions. One manifestation is the anisotropy induced by the presence of magnetic fields in the microscopic velocity distribution, which leads to measurable polarization effects (see Sect.~\ref{sec:pSZ}). Another is the non-thermal pressure support induced by turbulent bulk motions in the medium, measurable via the kSZ effect (Sect.\ref{sec:kSZveloc}). A third, very important situation can occur when the electrons (again at the microscopic level) have a relativistic non-thermal velocity distribution with extended high-momentum tails, resulting in an ``SZ increment" that can extend from the microwave domain all the way up to hard X-rays or gamma-rays. One generally refers to this last case as the ``non-thermal SZ'' (ntSZ) effect, which has a long history of discussion in the literature \cite{Rephaeli95,Enss00,Colafrancesco03,Colafrancesco13} but no definitive observational confirmation to-date.

\begin{figure}[bthp!]
\centering
\includegraphics[width=0.5\textwidth]{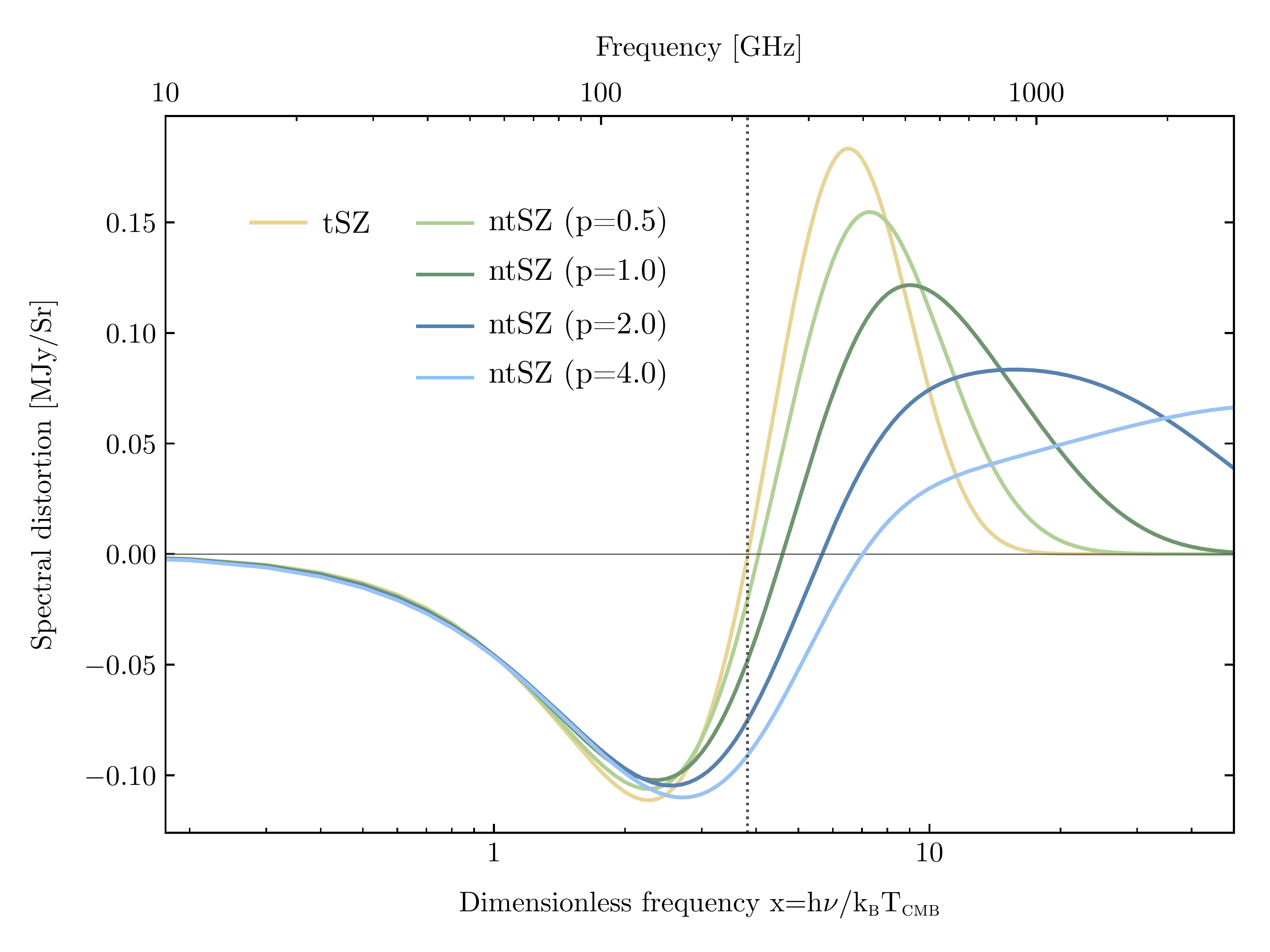}
\caption{\small Comparison of the non-relativistic tSZ spectrum and the non-thermal SZ spectrum for monoenergetic electrons with varying momentum, where $p$ is the dimensionless electron momentum (from \cite{Mroczkowski19}). The $y$- parameter for the tSZ case was set to $y = 10^{-4}$. We scaled the optical depth for the ntSZ to mimic a fixed overall $y$-parameter. The ntSZ contribution is expected to be a small fraction ($\approx 0.1$--1\%) of the tSZ signal in a galaxy cluster.}
\label{fig:ntSZ} 
\end{figure}

In Fig.~\ref{fig:ntSZ}, we show the non-thermal SZ (ntSZ) spectrum for a single-momentum electron distribution. The CMB photons are strongly up-scattered towards high frequencies once the momentum (expressed in units of $m_{\rm e} c$) exceeds $p \simeq 1$ \citep{Enss00, Colafrancesco03}. The ntSZ signal in a typical galaxy cluster is expected to be less than 1\% of the tSZ signal due to the overwhelming dominance of thermal pressure \cite{Zandanel14,Pinzke17}. 
However, this ratio can increase by a factor of several in radio-halo clusters  \cite{Brunetti14,vanWeeren19}, or even exceed 100\% within specific regions inside galaxy clusters, such as radio-AGN bubbles \cite{Pfrommer05,Abdulla18}.  The appeal of the ntSZ effect, compared to inverse-Compton measurements in X-rays (assuming the same electron population is responsible for both), is that ntSZ is sensitive to a far lower minimum-energy threshold of the cosmic-ray electrons \cite{Colafrancesco13}, and hence it is more suitable for determining the total relativistic energy budget. 
The ntSZ effect can also allow us to shed new light on the nature of dark matter and annihilating particles \cite{Colafrancesco04}.

%--------------
%--------------

To measure the ntSZ effect it is crucial to have sufficient high-frequency coverage. 
The ``zero crossing" of the ntSZ spectrum can extend to very high frequencies in the submillimeter ($\nu\simeq 0.3$--1\,THz), and only at higher frequencies one can witness the reappearance of the scattered photons after the strong energy exchange with the non-thermal relativistic electrons. Even the rSZ effect is usually sub-dominant in this regime. Measurement of the ntSZ effect will be a fundamental step forward in determining the total cosmic-ray energy budget of the Universe, 
and hence the strength of the magnetic field on large scales.  
This will form unique synergy with low-frequency experiments like the Square Kilometre Array, which will measure the same objects in radio synchrotron emission. This aspect becomes even more relevant as no upcoming or planned X-ray mission in the next decades will have the requisite imaging capability in the hard X-ray range to measure this inverse-Compton signal from galaxy clusters and cosmic filaments.  
%Measurements of the ntSZ effect could thus provide a unique probe of the cosmic ray energy budget in the Universe. 
The ntSZ effect will also lead to an all-sky non-thermal spectral distortions that can be targeted using absolute CMB spectroscopy without the necessity to resolve individual sources \citep{Chluba2019WP}.

\subsubsection{Resonant scattering and other frequency-dependent CMB signals} 
\label{sec:rescatt}

Resonant scattering of CMB photons produces spectral and spatial distortions that can be used to obtain tomographic information of the Universe, from recombination to the pre-reionization epoch \cite{Basuetal2004,HS05}. These signals will be accessible for the first time with \Backlight, and will provide an alternative and complementary window into the epoch of recombination, the Dark Ages, and the era of cosmological reionization. Varying observing frequencies will allow us to conduct {\sc Hi} tomography during the epoch of dynamical decoupling between baryons and photons and cosmological recombination. Moreover, while {\sc Hi} 21\,cm observations will probe the neutral Universe, CMB observations will map the metal-polluted Universe, usually associated to the ionized, star-forming regions of early epochs, thus providing a complementary view of this elusive era of our Universe. \\

\begin{figure}[tbh]
\centering
\includegraphics[width=0.5\textwidth]{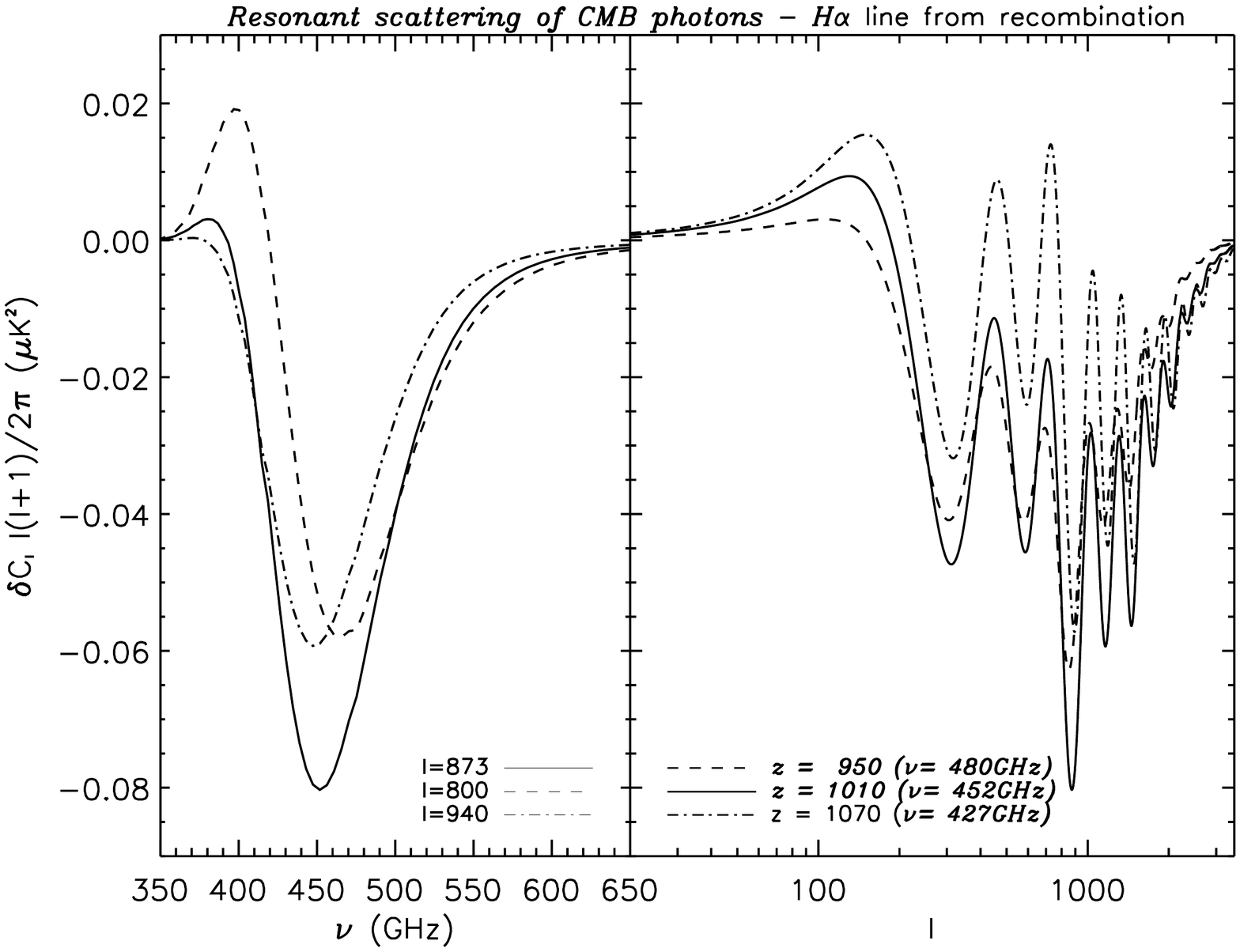}~
\includegraphics[width=0.56\textwidth]{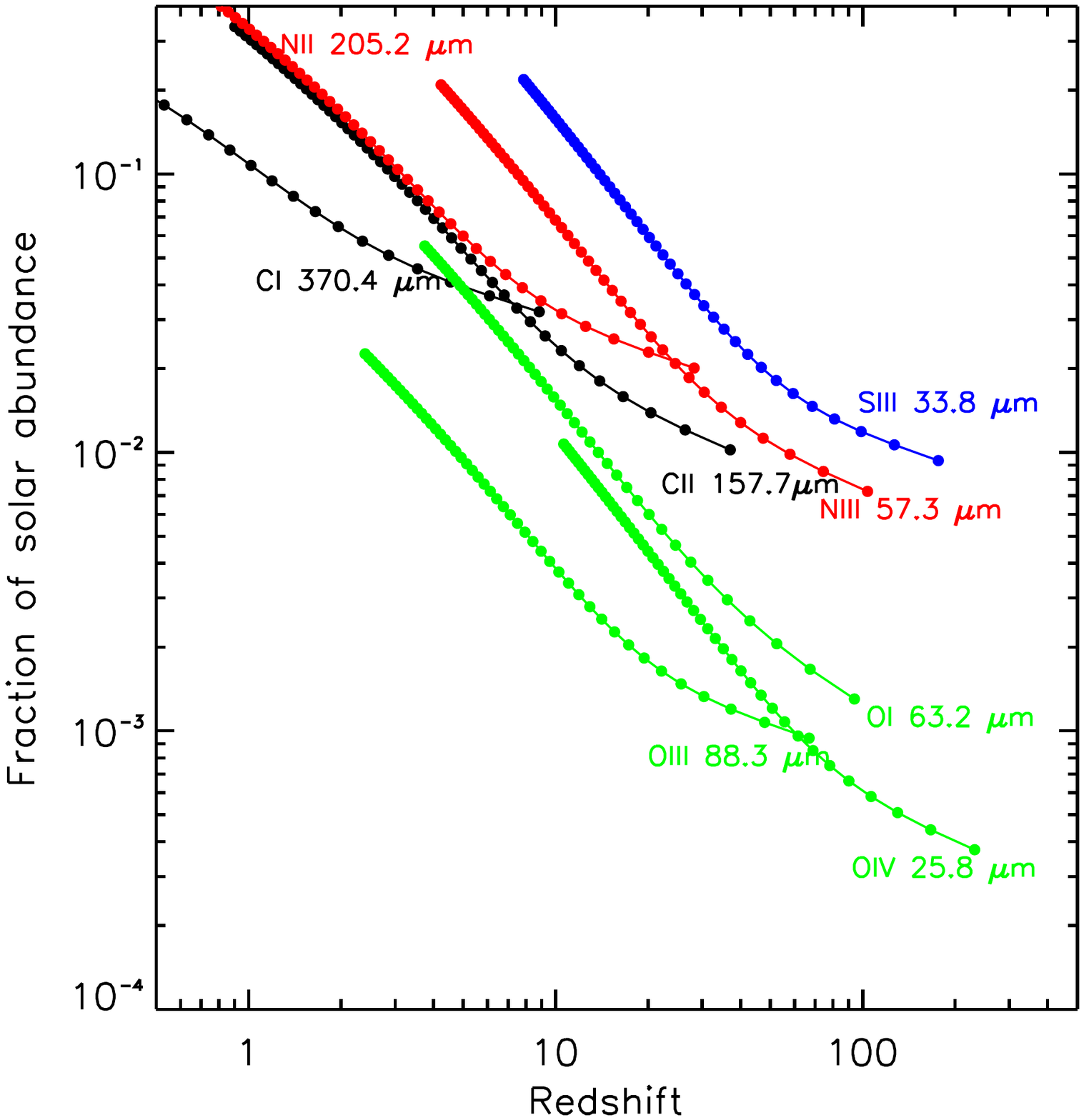}
\caption{\small
\textit{Left}: Relative change in the CMB angular power spectrum ($TT$) arising from the resonant scattering on the H$\alpha$ line generated during recombination, as a function of the redshifted frequency (left panel) and the angular multipole (right panel). Solid lines in both panels refer to the cases in which the signal is largest.  This is adapted from Ref.~\cite{RHS05}. \textit{Right}: Constraints at the 3$\sigma$ level on the abundance of metals inferred from different fine-structure transitions at different redshifts for a multi-frequency CMB mission covering the frequency range $50$--$1000$\,GHz. It is assumed here that the inter-channel calibration uncertainty lies at the 0.001\,\% level, and these constraints scale linearly ($\propto f_{\rm cal}/10^{-5}$) with respect to this calibration uncertainty parameter. The impact of other foregrounds is less critical, since their angular pattern differs from that of the CMB \citep{Hernandezetal2006}.}
\label{fig:Halpha} 
\end{figure}

{\bf Hydrogen line-scattering during recombination:} 
Resonant scattering of CMB photons in the hydrogen lines during cosmological recombination introduces a frequency-dependent optical depth contribution and hence a CMB anisotropy signal \cite{RHS05, Hernandezetal2007a}. The effect on the $TT$ power spectrum from the first lines of the Balmer and Paschen series will be detectable with the proposed \Backlight, providing an additional opportunity to directly constrain the recombination history and also to obtain independent determinations of the cosmological parameters (e.g., $\Omega_{\rm b}$ or $\Omega_{\rm m}$). 
The left panel of Fig.~\ref{fig:Halpha} shows the prediction for the H$\alpha$ line, with a maximum amplitude of around 0.3\,$\mu$K at frequencies around 450\,GHz and at angular scales corresponding to $\ell \approx 870$. For the P$\alpha$ line, the intensity reaches $0.02$\,$\mu$K at 160\,GHz and scales of $\ell \approx 890$ \cite{RHS05}. 
Those spectral features should also be detectable in linear polarization ($TE$ and $EE$ spectra), at least for the  H$\alpha$ line. For example, the maximum amplitude of the $EE$ signal for the H$\alpha$ line is $0.05$\,$\mu$K at frequencies of 430\,GHz and angular scales of $\ell \approx 1000$ \cite{Hernandezetal2007a}. \\

%--------------
%--------------

{\bf Metals during the dark ages:}
Any scattering of CMB photons after recombination blurs the primordial CMB anisotropies at small angular scales, while producing new anisotropies at intermediate angular scales that depend on the redshift of the scattering event.
The resonant scattering of CMB photons by fine-structure lines of metals and heavy ions produced by the first stars constitutes a frequency-dependent optical depth \cite{Basuetal2004}. Similarly to the case of cosmological hydrogen recombination, by comparing the CMB temperature and polarization anisotropies at different frequencies, one can thus probe the abundances of ions such as {\sc Oi}, {\sc Oiii}, {\sc Nii}, {\sc Niii}, {\sc Ci}, and {\sc Cii} at different redshifts \citep{Hernandezetal2006,Hernandezetal2007a}. 
 On intermediate angular scales ($\ell=50$--5000), the dominant effect that arises when comparing two different observing frequencies $i$ and $j$ is the blurring of the original CMB anisotropy field, $\delta C^{ij}_\ell \simeq -2 \tau_{ij} C_\ell^{\rm CMB}$, where $\tau_{ij}$ stands for the effective increase of the optical depth on all existing resonant lines from frequency $i$ to $j$. %In Fig.~\ref{fig:metals} 
 In the right panel of Fig.~\ref{fig:Halpha} we display the constraints on the abundance of different metal species that could be attained from \Backlight. At the map level, it turns out that, on intermediate angular scales, $\delta a_{{\ell}m} = - \tau_{ij} a_{{\ell}m}^{\rm CMB}$, with $a_{{\ell}m}^{\rm CMB}$ being the initial, intrinsic CMB anisotropy field. This fact should enable  easier separation of the metal-induced component from other frequency dependent signals \citep{Hernandezetal2007a}.\\

{\bf Collisional emission of metals during the Dark Ages:}
On small scales, the collisional emission associated to fine-structure transitions of metals and molecules addressed above in the context of resonant scattering will introduce dominant spectral and angular distortions in the CMB. The amplitude of the angular anisotropies will depend not only on the abundance and thermal state of the metals, but also on spectral resolution of the CMB observations \citep{metals_17}. 
At the same time, UV radiation emitted by the first stars can push the {\sc Oi} 63.2\,$\mu$m and {\sc Cii} 157.7\,$\mu$m transitions out of equilibrium with the CMB, producing a distortion $\Delta I_\nu/I_\nu \simeq 10^{-8}$--$10^{-9}$ due to fine-structure emission \cite{Gongetal2012,Hernandezetal2007b}, providing yet another window to reionization which should be potentially observable with \Backlight. \\

{\bf Rayleigh scattering:} Rayleigh scattering on neutral hydrogen during and after recombination induces another frequency-dependent signal ($\propto \nu_{\rm obs}^{4}$), both in intensity and polarization, that can provide further insight into the early Universe \citep{YuSpergelOstriker2001, Lewis13}. \Backlight\ will measure the temperature Rayleigh cross-spectrum with sub-percent accuracy,  detect the polarization from Rayleigh scattering also at high significance, and trace the cross-spectra between the Rayleigh temperature signal and primary polarization.

%%%%%%%%%%%%%%%%%%%%%%%%%%%%%%%%%%%%%%%%%%%%%%%%%%%%%%%%%%%%%%%%
\vspace{2mm}
\section{Complementarity of Space- and Ground-based Experiments} 

\begin{figure}[tbh]
\centering
%\begin{center}
%  \includegraphics[width=\textwidth, height=7cm, clip=true, trim=0 150 0 140]{Figures/ground_vs_space.pdf}
  \includegraphics[width=0.5\textwidth, clip=true, trim=0 -5 0 0]{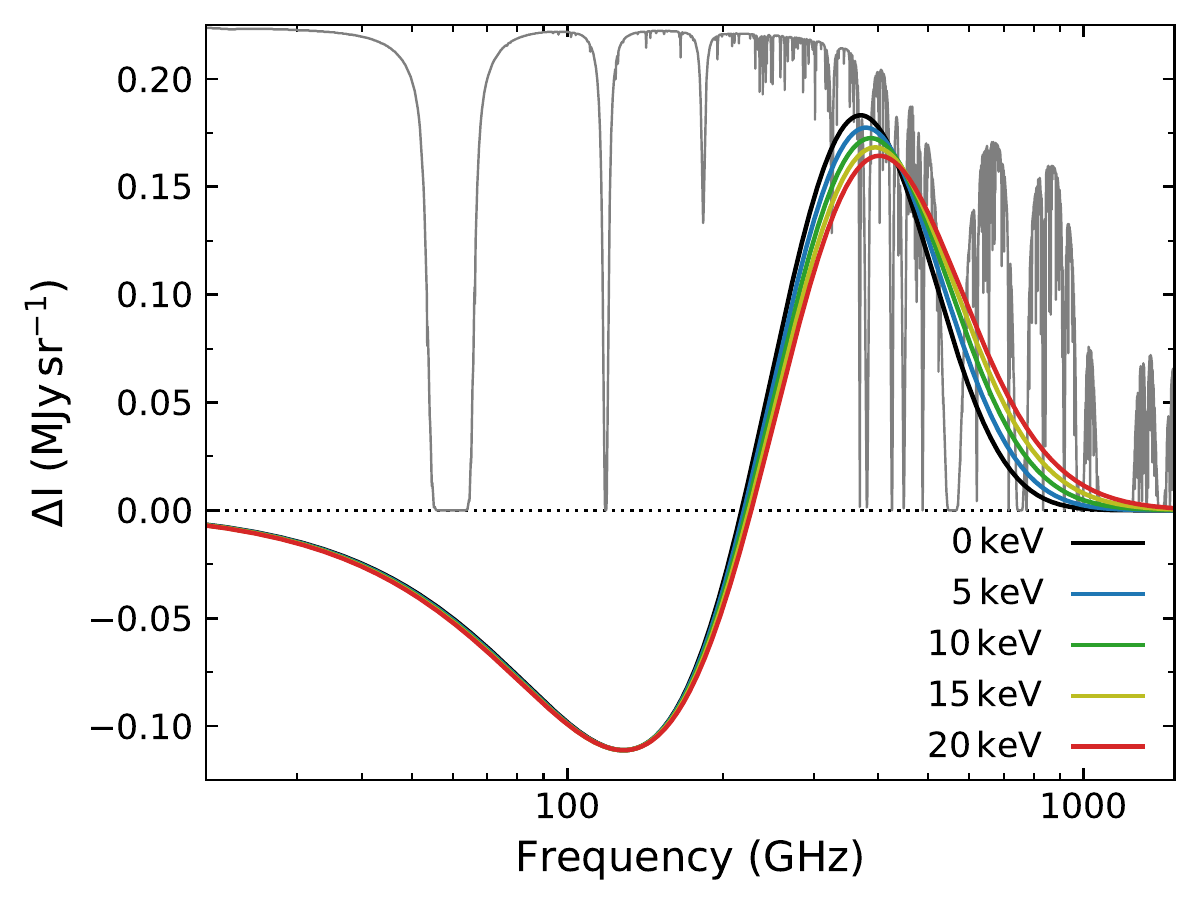}~
\includegraphics[width=0.5\textwidth, height=6.7cm]{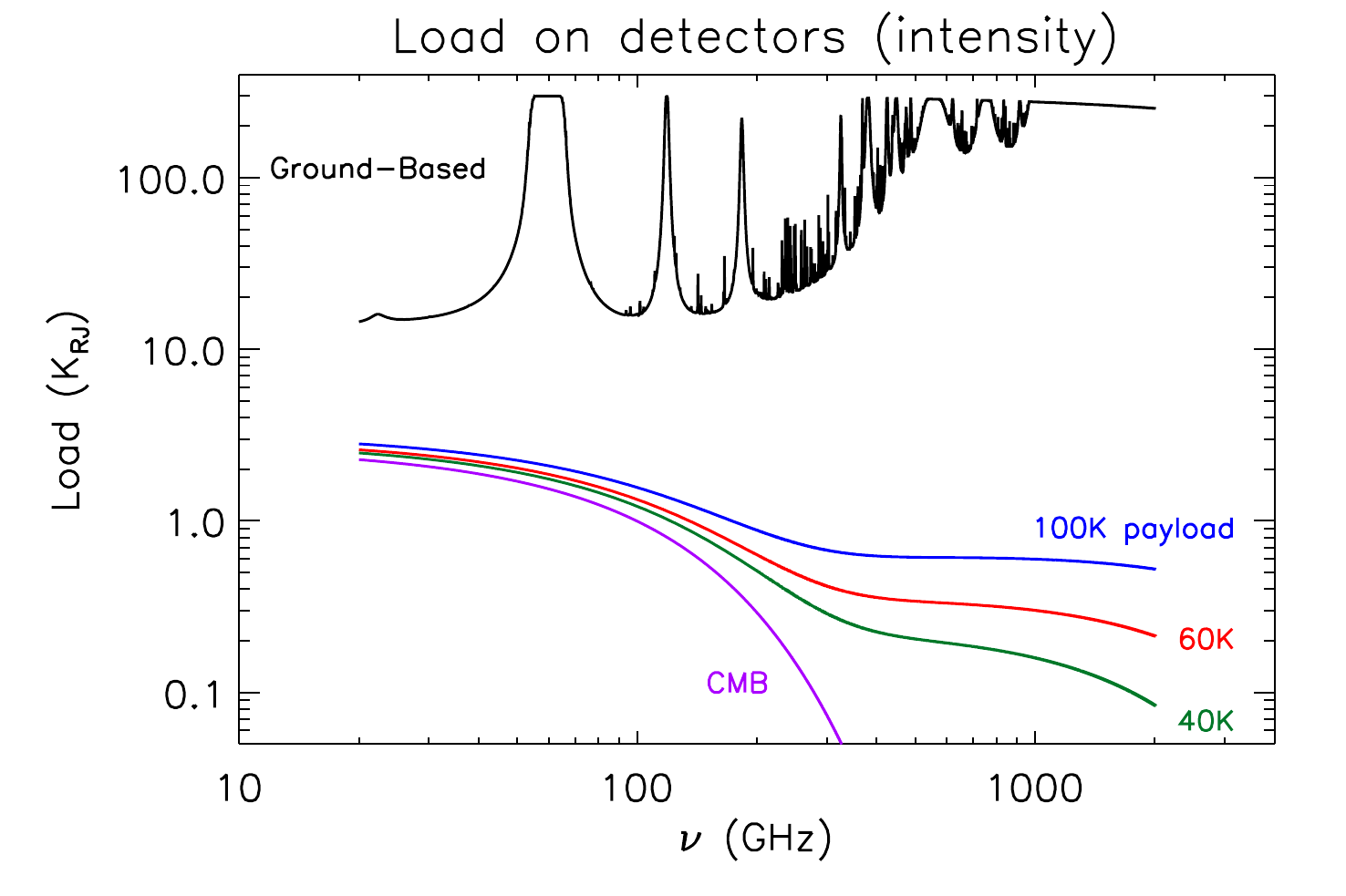}
%\end{center}
\caption{\small {\it Left:} Spectrum of the thermal SZ effect with relativistic corrections (rSZ effect), with a varying $y$-parameter designed to match the spectra at the minima of the tSZ decrement. On top we show the atmospheric transmission (scaling from 0 to 1) from one of the best sites on Earth, on Cerro Chajnantor in the Atacama.
{\it Right:} Load on detectors in space versus those on the ground (from Ref.~\cite{COREmission}). 
The rapidly worsening atmospheric load from the ground in the submillimeter regime is one of the prime motivations for going to space and this advantage is further enhanced by having a cold primary mirror. 
}
\label{fig:ground-space}
\end{figure}

The main complementarity of ground-based and space-based CMB experiments concerns {\it spectral coverage}, {\it sky coverage}, and {\it spatial resolution}. Current, planned, and future concepts for ground-based CMB experiments such as SPT \cite{Carlstrom2011}, ACT \cite{Fowler2007}, BICEP/Keck, Class, Simons Observatory \cite{SimonsObs2019}, and CMB-S4 \cite{CMBS42016, S4DSR19} are constrained to observe a fraction of the sky at low frequencies in a handful of atmospheric windows spanning the 30--270\,GHz range with limited results from the 220--270\,GHz bands, while space missions can observe the entire sky at high frequencies $>$ 300\,GHz without the significant challenges presented by the atmosphere at those frequencies. %({\it Notable exceptions from the ground include CCAT-prime, now under construction, and potentially AtLAST, both of which include first-light designs or concepts surveying up to $\sim$950~GHz}).
Ground-based studies offer the potential for building much larger telescope infrastructures, featuring apertures currently challenging to construct in space, hence providing high spatial resolution in the low-frequency range 30--270\,GHz (however, see \cite{Mukherjee2019} for in-space telescope assembly). 
In contrast, space missions are limited to relatively low spatial resolution at low frequencies but can offer spatial resolution in the high-frequency range 300\,GHz--1\,THz similar to what is achievable from the ground at 30--270\,GHz. Therefore, a combination of ground-based and space-based observations would provide comprehensive spectral window of 30\,GHz--1\,THz covering the full SZ spectrum (Fig.~\ref{fig:ground-space}; \textit{left}) with quite uniform high spatial resolution. Finally, while ground-based observatories will detect cluster samples and provide maps of individual clusters over portions of the sky, space missions can deliver all-sky legacy cluster surveys and map the diffuse SZ emission at large angular scales for cosmological studies.

Ground-based facilities benefit from accessibility and less stringent power requirements, allowing upgraded instrumentation every few years and much larger detector counts readout by back-end electronics that can have both high data rates and high power consumption. The accessibility of ground-based experiments allows for detectors with lower technical readiness levels (TRLs) to be used, as each season's debugging and maintenance can lead to substantial progress and improvements.  In comparison, space-borne missions necessarily feature fewer detectors, but at an extremely high TRL, and with readout electronics that consume less power, can be cooled passively, and can reliably transmit the data back to Earth. The most salient difference between space and ground-based experiments is, of course, atmospheric transmission \cite{Radford16, Otarola2019}, which limits the spectral windows useful for ground-based experiments.  In addition, the emission from the atmosphere and time-varying fluctuations and the entrance window both contribute a lot of photon noise (see Fig.\,3 of \cite{Choi2019}), requiring several orders of magnitude more detectors from the ground to obtain a comparable mapping speed as from space (see Figs.~11 and 12 in Ref.~\cite{COREmission}).

All-sky maps at frequencies $>$ 300\,GHz delivered by past space-mission surveys like ESA's \textit{Planck} have forged an important legacy to the Astronomy community for decades. They have been of major importance for most ground-based CMB experiments to characterise residual foreground contamination in their data and to obtain evidence for false detections. As a historical example, the  \textit{Planck} 353-GHz map has served as an exquisite tracer of the Galactic thermal dust contamination in the CMB $B$-mode data of the ground-based experiment BICEP2, providing evidence for {\it false} detection of the primordial gravitational wave signal from inflation by BICEP2 \citep{BICEP2-Planck2015}. Dust and CIB foregrounds are extremely challenging to characterise at frequencies $<270$\,GHz by ground-based CMB surveys, which often have to rely on the extrapolation of the high-frequency templates provided by space-mission surveys. \\

{\bf A new era of faint signal-to-foreground regimes:} The next decades will be dedicated to the search for ever fainter cosmological signals, e.g., kSZ, rSZ, pSZ, and CMB-cluster lensing as described in this white paper, for which high-frequency observations at high resolution from \Backlight\ of much higher precision than \textit{Planck} will be of crucial importance to control unavoidable foreground biases. Besides the control of foregrounds, spectral coverage at high frequencies $>300$\,GHz allowed by the space mission is essential to discern the distinct spectral signature of some of the signals presented here, such as rSZ and non-thermal SZ effects. Finally, having an absolute spectrometer on board as an option for absolute calibration \citep{Chluba2019WP} would be a huge gain for measuring these faint SZ signals without bias.

%%%%%%%%%%%%%%%%%%%%%%%%%%%%%%%%%%%%%%%%%%%%%%%%%%%%%%%%%%%%%%%%
\vspace{2mm}

\section{Requirements for a Space Mission}
\label{sec:spacesim}

{\bf Mission characteristics:} Successful application of the techniques  described in Sect.~\ref{sec:scicase} demands  highly accurate separation of numerous astrophysical signals with differing spectra. This requires at a minimum an \emph{imager observing in multiple frequency channels over the range from 50\,GHz to 1\,THz}. \emph{At least 20 frequency channels} are needed to disentangle the target signals from contamination by sources of foreground and background emission, and to separate the different signals from the studied structures themselves (tSZ, rSZ, kSZ, pSZ, non thermal SZ, infrared and radio sources). A frequency range of 50\,GHz to 1\,THz insures full coverage of the SZ spectra and also accurate modeling of dust spectral energy distributions and cosmic infrared background correlations across frequencies.

Observation of the faint signals from  filaments and low mass ($M\sim 5 \times 10^{13} {\rm M}_\odot$) halos requires an average \emph{sensitivity of order a few times 0.1\,$\mu {\rm K}$-arcmin}, at least over the channels between 100 and 250\,GHz. The \emph{imager must be polarization sensitive} to detect the polarized SZ effects and to monitor expected systematic uncertainties in the measurement of halo lensing with CMB intensity.

To resolve cluster sized halos, we need  \emph{a survey with a 1$^\prime$ (goal) to 1.5$^\prime$ (requirement) beam} (the angular radius $\theta_{500}$ of a $10^{14}{\rm M}_\odot$ cluster at $z=1$ is 1$^\prime$). This resolution can be achieved from the ground in atmospheric windows below 300\,GHz, but \emph{outside these windows and above 300\,GHz, it can only be achieved from space} and requires a cold 3- to 4-m class telescope, or preferentially one of 4- to 6-m class. 

We emphasize that \Backlight\ can attain its science goals alone, without ground-based millimeter data; nevertheless, this science would greatly benefit from the combination of \Backlight\ and the CMB-S4\ experiment, the latter providing complementary deep, arcmin-resolution maps at frequencies below 300\,GHz.

{\bf Mission class:} The need for a cold 3 to 4m class telescope (preferentially, 4 to 6m class) places the mission in the \emph{L class category}. Decreasing the survey resolution to worse than 1.5$^\prime$ at 300\,GHz may fit within the envelope of a M class mission, but it would not allow to attain the science goals described in this white paper. Furthermore, decreasing the survey resolution would lose the good match   between the space survey and the future CMB-S4 ground based survey, largely weakening the promising combination of the two data sets. We thus require to keep the resolution better than 1.5$^\prime$ at 300\,GHz for the \mission. 
%% added sentence on in-space assembly
We also note that in the future, the scope and derived costs of space observatories that carry large mirrors, cooling systems or other similar hardware might benefit significantly by taking advantage of modular designs and {\it in-space assembly}, as has been shown by recent studies \cite{Mukherjee2019}.

{\bf Synergy with ground based experiments:} As stated above, \Backlight\ can explore all the science cases described in this white paper without the need from external millimeter data. However \emph{the proposed concept is in perfect synergy with the planned ground based experiment CMB-S4}, which will perform a survey at the same resolution 1.5$^\prime$ and equivalent depth, in atmospheric windows at frequencies below 300\,GHz. Very high resolution millimeter projects such as CMB-HD \cite{Sehgal2019} and AtLAST \cite{Klaassen2019} would complement the survey from the \mission\ in providing even higher resolution maps of specific fields below 300\,GHz.

\vspace{2mm}
\section{Conclusion}

Following the rapid progress in cosmology and extrapolating it over the coming decade, we can anticipate that a full picture of structure formation and evolution will still be lacking by the mid-2030s. To make a very significant breakthrough in this regard, we propose to use the cosmic microwave background  as a ``backlight'' to probe structures across the entire observable Universe. Our proposed Backlight Mission would use gravitational lensing, various types of Sunyaev-Zeldovich effects, as well as resonant and Rayleigh scattering to: (1) trace the distributions of both dark matter and baryons; (2) explore the different physical state of the baryons; and (3) measure the cosmic velocity field. This mission would achieve, for the first time, the long-sought goal of a complete census of the total mass, gas, and stellar content of the Universe, giving a clear picture of the cosmic web and its evolution. It will also enable a clearer insight into the nature of dark energy and the ultimate fate of the Universe. 
Our goal is to envision a space mission whose rich legacy dataset will form the ideal complement to many terrestrial experiments that are currently operational or proposed for the future, setting the stage for a host of groundbreaking discoveries for decades to come. %

%%%%%%%%%%%%%%%%%%%%%%%%%%%%%%%%%%%%%%%%%%%%%%%%%%%%%%%%%%%%%%%%%
\pagebreak
%\textbf{References}

\bibliographystyle{aasjournal}

\bibliography{szrefs,Nagai,Naonori,Churazov,chmrefs,resonant,Basu,Chluba}

\end{document}